\documentclass[structabstract,dvipsnames]{aa}
\pdfoutput=1

\ifx\pdfoutput\undefined
\usepackage{graphicx}
\else
\usepackage[pdftex]{graphicx}
\usepackage[pdftex,pdfauthor={J.~W.~Threlfall},citecolor=blue,linkcolor=blue, linktocpage=true,colorlinks=true]{hyperref}
\usepackage{epstopdf}
\fi 

\usepackage{natbib}
\bibpunct{(}{)}{;}{a}{}{,} 
\usepackage{txfonts}
\usepackage{graphicx}	
\usepackage[usenames]{color}
\usepackage{amssymb}	
\usepackage{aas_macros}	
\usepackage{subfig}	
\usepackage{fixltx2e}	
\usepackage{mathrsfs}
\usepackage{tabularx}
\usepackage{hypcap}

\newcommand{\beq}{ \begin{eqnarray} }
\newcommand{\eeq}{ \end{eqnarray} }

\newcommand{\zhat}{ {{\bf{\hat z}}} }

\newcommand{\dive }{ {\bf {\nabla}} \cdot }
\newcommand{\curl }{ {\bf {\nabla}} \times }

\newcommand{\boldnabla}{\mbox{\boldmath$\nabla$}}

\def\A{{Alfv\'en}}

\def\ca{{c_A}}

\def\grad{\boldnabla}

\def\muo{\mu_{0}}

\def\zhat{\bf{\hat{z}}}

\def\di{\delta_i}

\date{}			

\begin{document}
\title{{Nonlinear wave propagation and reconnection at magnetic X-points in the Hall MHD regime}}
\author{J. W. Threlfall\inst{\ref{inst1}} \and C. E. Parnell\inst{\ref{inst1}} \and I. De Moortel\inst{\ref{inst1}} \and K. G. McClements\inst{\ref{inst2}} \and T. D. Arber\inst{\ref{inst3}}}
\institute{School of Mathematics and Statistics, University of St Andrews, St Andrews, Fife, KY16 9SS, U.K. \email{jamest@mcs.st-and.ac.uk;clare@mcs.st-and.ac.uk;ineke@mcs.st-and.ac.uk}\label{inst1}
\and United Kingdom Atomic Energy Authority, Culham Science Centre, Abingdon, Oxfordshire, OX14 3DB, U.K.
\email{k.g.mcclements@ccfe.ac.uk}\label{inst2}
\and Centre for Fusion, Space and Astrophysics, Department of Physics, University of Warwick, Coventry, CV4 7AL, U.K.
\email{T.D.Arber@warwick.ac.uk}\label{inst3}}

\abstract
{The highly dynamical, complex nature of the solar atmosphere naturally implies the presence of waves in a topologically varied magnetic environment. Here, the interaction of waves with topological features such as null points is inevitable and potentially important for energetics. The low resistivity of the solar coronal plasma implies that non-MHD effects should be considered in studies of magnetic energy release in this environment.}
{This paper investigates the role of the Hall term in the propagation and dissipation of waves, their interaction with 2D magnetic X-points and the nature of the resulting reconnection.}
{A Lagrangian remap shock-capturing code (Lare2d) is used to study the evolution of an initial fast magnetoacoustic wave annulus for a range of values of the ion skin depth ($\delta_i$) in resistive Hall MHD. A magnetic null-point finding algorithm is also used to locate and track the evolution of the multiple null-points that are formed in the system.}
{Depending on the ratio of ion skin depth to system size, our model demonstrates that Hall effects can play a key role in
the wave-null interaction. In particular, the initial fast-wave pulse now consists of whistler and ion-cyclotron components; the dispersive nature of the whistler wave leads to (i) earlier interaction with the null, (ii) the creation of multiple additional, transient nulls and, hence, an increased number of energy release sites. In the Hall regime, the relevant timescales (such as the onset of reconnection and the period of the oscillatory relaxation) of the system are reduced
significantly, and the reconnection rate is enhanced.}
{}
\maketitle

\section{Introduction}\label{sec:Intro}

Observations of the Sun and theoretical models have demonstrated that complicated geometrical and topological magnetic structures pervade the solar atmosphere. Furthermore, the ever-improving cadence and sensitivity of instruments continue to reveal its highly dynamic nature through the presence of many types of waves \citep[see e.g.][]{review:DeMoortel2005,review:NakariakovVerwichte2005,review:DeMoortelNakariakov2012}. It is therefore natural to expect that waves will interact with topological features, including magnetic null points; such interactions have formed the basis of many investigations in both 2D and 3D \citep[see][and references therein]{review:McLaughlinetal2011}. 

Null points, especially in 2D, are known to be important sites of energy release, as first demonstrated by \citet{paper:Dungey1953}, \citet{paper:Parker1957} and \citet{paper:Sweet1958}, although in 3D they are not the only sites where reconnection may occur \citep[initially demonstrated by][]{paper:Schindleretal1988,paper:HesseSchindler1988}; see also \citet{book:Biskamp} and \citet{book:PriestForbes} for reviews of the considerable body of work that focuses on this process. Null point reconnection is suspected to play a major role in certain types of CMEs \citep[see e.g. the magnetic breakout models of][]{paper:Antiochosetal1999,paper:Pariatetal2009} and solar flares \citep[see e.g.][]{paper:Fletcheretal2007,paper:Massonetal2009,paper:Massonetal2012}.

It is estimated that there are more than $20,000$ null points at any one time during solar minimum in the solar corona above $1.5$Mm \citep{paper:LongcopeParnell2009}. Below this height, the techniques used to estimate the number of nulls are unreliable, however, since the complexity of the magnetic field is known to fall off rapidly with height, it is anticipated that there will be considerably more magnetic null points below this height than above.

Assuming the electrical resistivity of solar coronal plasma is due to electron-ion collisions, the low collisionality of such a plasma implies that significant energy release can only occur through resistive dissipation if the current density is extremely high and the magnetic field scale length is very small, thereby calling into question the self-consistency of resistive magnetohydrodynamic (MHD) reconnection models. Moreover, resistive MHD generally yields reconnection rates that are too low to account for the observed timescales of energy release in solar flares \citep[see e.g.][]{paper:CraigMcClymont1991}.

Both theoretical considerations and observations thus motivate us to construct models of reconnection that go beyond resistive MHD. One relatively straightforward way of doing this is to include the Hall term in Ohm's law. This term does not by itself give rise to reconnection, but it can accelerate the rate of reconnection in the presence of finite resistivity \citep{paper:Birnetal2001,book:BirnPriest}. Generally, Hall effects are expected to become significant when length scales approach the ion skin depth, $\delta_i$ and wave frequencies approach the ion cyclotron frequency. In view of the low resistivity and consequently small resistive length scale of the coronal plasma, noted above, it is natural to include such effects in models of reconnection (for example, in solar flares), despite the fact that the reconnection length scales cannot be resolved observationally. When the plasma beta is less than unity, as in the upper chromosphere and corona, $\delta_i$ is larger than the thermal ion Larmor radius, and Hall MHD is then applicable. Analytical Hall MHD models of steady incompressible reconnection \citep{paper:Dorelli2003,paper:CraigWatson2003} have been used to determine the conditions in which Hall currents can influence reconnection and Ohmic dissipation rates. More recently, \citet{paper:SenanayakeCraig2006} used the linearised Hall MHD equations to investigated numerically the energy decay of a perturbed line-tied X-point system. \citet{paper:CraigLitvinenko2008} extended this linear model to obtain more detailed reconnection scalings, finding, unsurprisingly, that Hall effects on the reconnection rate diminish as the system size is increased. 

The magnetic topology, and in particular null points, have also been shown to affect wave propagation in their vicinity \citep[summarised in][]{review:McLaughlinetal2011}. Fast magnetoacoustic waves tend to "wrap around" magnetic null points \citep{paper:McLaughlinHood2004}, and progressively decreasing length scales in the vicinity of the null can give rise to nonlinear effects and drive reconnection \citep{paper:McLaughlinetal2009}.

Many studies of both reconnection and wave propagation in the solar context have employed a two-dimensional current-free magnetic X-point equilibrium with a single null \citep{paper:CraigMcClymont1991,paper:CraigMcClymont1993,paper:CraigWatson1992,paper:Hassam1992,paper:Ofmanetal1993,paper:McClementsetal2004,paper:McLaughlinHood2004,paper:McLaughlinHood2005,paper:McLaughlinHood2006b,paper:SenanayakeCraig2006,paper:CraigLitvinenko2008}. Although this particular equilibrium configuration is somewhat idealised, it has certain merits from a theoretical point of view. In addition to being current-free it is also scale free, and is thus characterised by no free parameters except for the magnitude of the magnetic field at a specified distance from the null. As noted by \citet{paper:McClementsetal2004}, the variation of Alfv\'en speed in a two-dimensional X-point configuration guarantees that a disturbance propagating towards the null will steepen as it does so, until it is modified by resistivity, non-MHD (e.g. Hall) effects, or nonlinear effects. Such a configuration is thus useful for exploring physical processes
related to energy release in 2D. Furthermore, understanding non-MHD processes in this 2D scenario is an essential first step before
considering more complex (3D) configurations.

The primary objective of the present work is to study numerically the nonlinear dynamics of a fast Alfv\'en disturbance propagating towards a two-dimensional X-point null, taking into account the Hall term in Ohm's law. Here, we specifically seek to address (i) how the Hall term manifests itself in models of wave-driven oscillatory reconnection and (ii) whether the rate of oscillatory reconnection is affected by the presence of the Hall term (as suggested by previous models of Hall MHD reconnection). In Sect.~\ref{sec:Setup} we outline our basic assumptions and equations, and the numerical setup used in simulations of fast waves propagating towards X-type magnetic nulls. The initial behaviour of the wave pulses is described in Sect.~\ref{sec:PulseBehaviour}, while the formation and evolution of complex magnetic topologies involving multiple null-points (in the case of simulations with finite $\delta_i$) is explained in Sect.~\ref{sec:Nulls}. For each value of $\delta_i$ the system ultimately settles settles into an oscillatory reconnective regime, described in Sect.~\ref{sec:Oscill}. A discussion of the nature of the reconnection process is presented in Sect.~\ref{sec:Reco}, followed by a summary and conclusions in Sect.~\ref{sec:Conc}.

\section{Numerical model setup}\label{sec:Setup}
The system was modelled numerically using a two dimensional version of a Lagrangian remap scheme ({\tt{LareXd}}), described by \cite{paper:LareXd2001}, which includes an optional Hall physics package to incorporate the Hall term into the standard MHD system of equations, seen here in normalised dimensionless form: 
\begin{eqnarray*}
&&\frac{\partial \rho}{\partial t}+\dive{\left(\rho{\bf{v}} \right) }=0, \\
&&\rho\left( \frac{\partial {\bf{v}}}{\partial t}+\left( {\bf{v}}\cdot\boldnabla\right){\bf{v}} \right)= \left(\curl{\bf{B}}\right)\times{\bf{B}}-\grad{p}, \\
&&\frac{\partial {\bf{B}}}{\partial t}=\curl{\left( {\bf{v}}\times{\bf{B}}\right) }-\curl{\left(\eta\curl{\bf{B}} \right) } -\lambda_i\curl{\left[\frac{1}{\rho}\left(\curl{\bf{B}}\right)\times{\bf{B}}   \right] }, \\
&&\rho\left( \frac{\partial\epsilon}{\partial t}+\left( {\bf{v}}\cdot{\boldnabla}\right)\epsilon\right)=-p\dive{\bf{v}}+\eta j^2,
\end{eqnarray*}
for dimensionless mass density $\rho$, pressure $p$, magnetic field strength ${\bf{B}}$, fluid velocity ${\bf{v}}$, internal energy density $\epsilon$, resistivity $\eta$ (the reciprocal of the Lundquist number) and ion skin depth $\lambda_i$. These equations have been normalised with respect to typical values of magnetic field strength ($B_0$), temperature ($T_0$) and number density ($n_0$). The velocities are scaled with the global {\A} speed $\ca\:(=B_0/\sqrt{\muo m_p n_0}$, with proton mass $m_p$ and permeability of free space $\muo)$, and time $t_0$ is scaled by the global {\A} time, $t_0=l_0/\ca$, for a typical system lengthscale $l_0$.

For example, adopting typical flaring coronal normalisation values of $B_0=100\mbox{G}$ and $n_0= 10^{16}\:\mbox{m}^{-3}$ implies velocities are normalised by $v_0=\ca\approx 2\:\mbox{Mms}^{-1}$. Adopting a typical flaring coronal temperature, $T_0=2\times10^6\:\mbox{K}$,  means that $\beta_0\approx0.007$ (where $\beta_0$ is the local plasma beta, $2\muo p_0/B_0^2$, defined at a radius $l_0$ from the origin). The role of the Hall term is quantified by the ion skin depth, $\di$. The normalised dimensionless ion skin depth used in the numerical scheme ($\lambda_i$) is given by:
\[
 \lambda_i=\frac{\di}{l_0}=\frac{c}{l_0\omega_{pi}},
\]
with the ion plasma frequency $\omega_{pi}\:(=\sqrt{n_0 e^2/m_p \epsilon_0}$ for electron charge $e$ and permittivity of free space $\epsilon_0)$ fixed through the number density $n_0$. A choice of $\lambda_i=0.0072$ implies a normalising lengthscale $l_0\approx0.3\:\mbox{km}$, while increasing $\lambda_i$ by a factor of $10$ reduces $l_0$ by the same amount ($l_0\approx32\:\mbox{m}$). 
Finally, the dimensionless plasma resistivity is determined using
\[
 \eta=\frac{\eta_0}{\muo l_0v_0},
\]
where $\eta_0$ is the electrical resistivity ($=1/\sigma$ for electrical conductivity $\sigma$). As pointed out by \citet{paper:CraigLitvinenko2002}, the levels of resistivity in the flaring corona may be as much as a factor of $10^6$ higher than the collisional value. Using this enhancement factor and the normalisation described above, we find $\eta=0.0005$ where $\lambda_i=0.0072$, while a smaller enhancement ($10^5$) is produced with an identical level of $\eta$ when $\lambda_i=0.072$. 

\subsection{Equilibrium and initial conditions}
\begin{figure}[t]
 \centering\capstart 
\resizebox{\hsize}{!}{\includegraphics{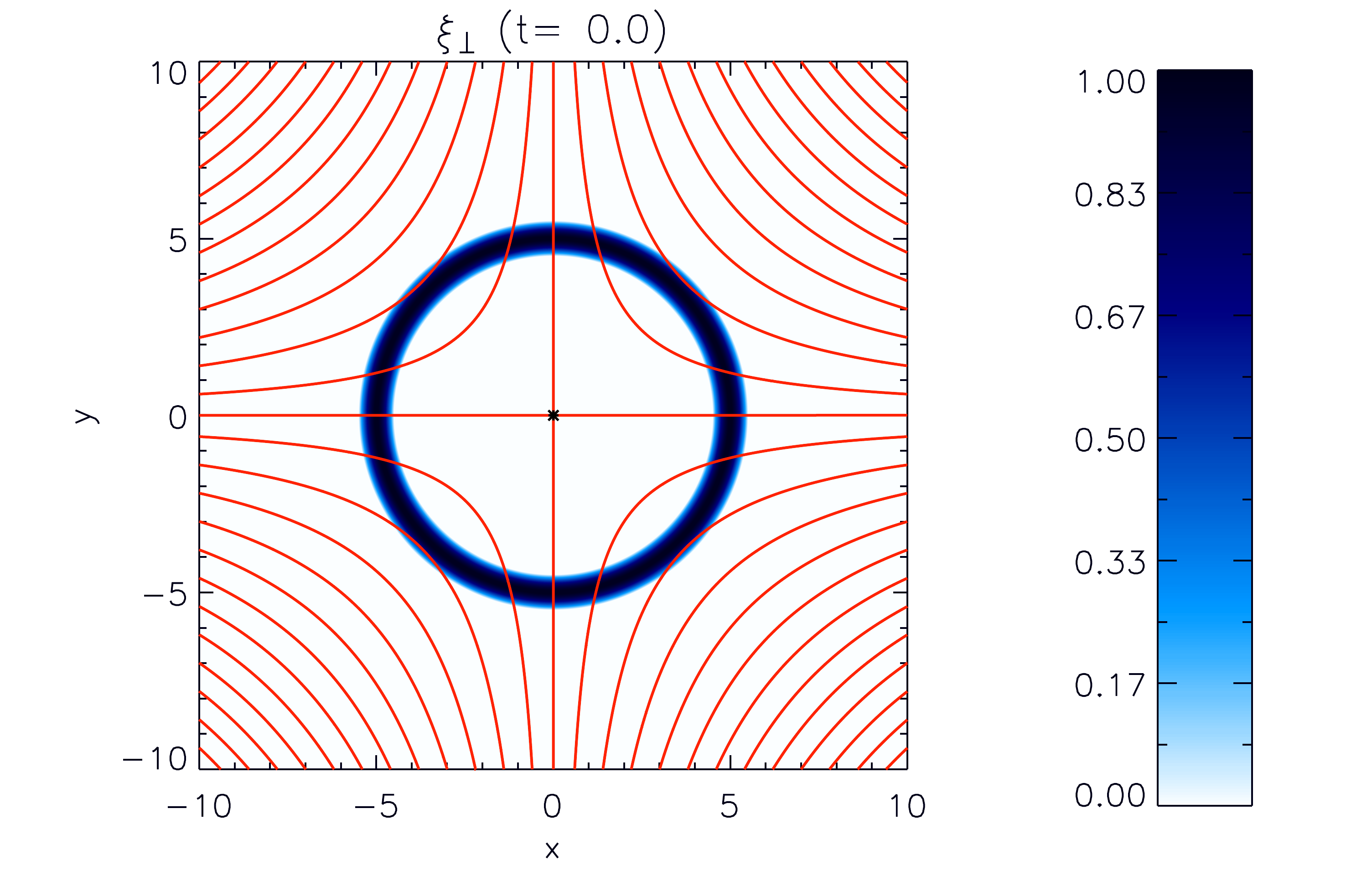}}
\caption{Initial fast wave annulus amplitude, defined using $\xi_\perp$, centred on $r=5$ with amplitude $A=1$. Also overplotted are selected contours of $A_z$ (seen in red), indicating the equilibrium field configuration ${\bf{B}}=\left[-x,y,0\right]$ with the X-point location indicated in white \citep[found using null-tracking routine of][]{paper:HaynesParnell2007}.} 
\label{fig:iniconditions}
\end{figure}

A wide range of equilibrium magnetic field configurations have been studied in investigations of wave/null-point interactions \citep[see][for a detailed review]{review:McLaughlinetal2011}. In this investigation, our equilibrium is chosen to be
\begin{equation}
 {\bf{B}}=\left[B_x,B_y,B_z\right]=\left[-x,y,0\right]. 
\label{eq:EqConfig}
\end{equation}
Note that this equilibrium configuration is curl-free (i.e. ${\bf{J}}=0$ initially) and contains a single null-point, located at the origin. A common tool for providing visual representation of 2D magnetic fields is the magnetic vector potential ${\bf{A}}$, which (for the initial field configuration detailed in Eq.~\ref{eq:EqConfig}) is given by
\[
 {\bf{A}}=[0,0,A_z]=[0,0,C-xy],
\]
where $C$ is an arbitrary constant. Lines of constant magnetic flux can be illustrated using contours of $A_z$, as seen in Fig.~\ref{fig:iniconditions}, however this figure illustrates only selected contours of $A_z$ over half of the simulated domain, $(x,y)\in[-10,10]$. 

A $5120\times5120$ grid was used to simulate a range $(x,y)\in[-20,20]$. Reflective (zero gradient) boundary conditions were applied to all variables (${\bf{B}},\rho,\epsilon$) except velocities, which were fixed to zero at all boundaries. {Early studies by \citet{paper:McLaughlinHood2004} showed that an initially planar MHD fast wave disturbance would refract around and accumulate at a magnetic null-point due to the decrease in {\A} speed. Once accumulated at the null, the wave would essentially become azimuthally symmetric, but would never actually reach the null (as $B\rightarrow0, \ca\rightarrow0$). By initialising an azimuthally symmetric disturbance with a large amplitude, \citet{paper:McLaughlinetal2009} studied how nonlinear effects would allow the wave to interact with the null. To aid direct comparison with \citet{paper:McLaughlinetal2009}, we retain their selection of natural system variables, 
\begin{eqnarray*}
\xi_{\perp}(x,y,t)&=&\left({\bf{v}}\times{{\bf{B}}}\right)\cdot{\zhat}=v_xB_y-v_yB_x, \\
\xi_{||}(x,y,t)&=&\left({\bf{v}\cdot{\bf{B}}}\right)=v_xB_x+v_yB_y,
\end{eqnarray*}
in order to specify similar initial conditions:
\begin{eqnarray*}
\xi_{\perp}(x,y,0)&=&A\sin{\left[\pi \left(r-4.5 \right)\right]  } \qquad 4.5\leq r\leq 5.5, \\
\xi_{||}(x,y,0)&=&0,
\end{eqnarray*}
where $r=\sqrt{x^2+y^2}$ (and also noting that velocities perpendicular and parallel to the local magnetic field may be found by scaling $\xi_\perp$ and $\xi_{||}$, i.e. $v_\perp=\xi_\perp/|\bf{B}|$ and $v_{||}=\xi_{||}/|\bf{B}|$ respectively). A visual representation of these initial conditions can also be seen in Fig.~\ref{fig:iniconditions}. 

In anticipation of the annulus splitting into two wave pulses (which travel radially inwards and outwards) and in order to focus only on the part of the pulse which interacts with the null, all system variables outside a radius $r=6$ were reset to their equilibrium values. This was performed at $t=0.7\tau_A$, the point at which the waves were deemed to be sufficiently independent from one another that the outer pulse may be removed without affecting the inner pulse. At this point, a damping layer was also introduced for $r\geq6$, removing kinetic energy from any waves which travel outwards beyond this radius, to avoid complications from reflection at the simulation boundaries.

\section{Fast wave annulus evolution in Hall MHD}\label{sec:PulseBehaviour}
As mentioned above, the initial fast wave annulus (shown in blue in Fig.~\ref{fig:iniconditions}) splits into two oppositely travelling wave pulses, travelling radially inwards and outwards, with each pulse initially having amplitude $0.5A$. Using the MHD case \citep[described in][]{paper:McLaughlinetal2009} as a benchmark, the effect of the Hall term on the splitting of the wave pulse is illustrated in the first column of Fig.~\ref{fig:annulusVperp}, for both the $\lambda_i=0.0072$ and $\lambda_i=0.072$ simulations.

\begin{figure*}[t]
  \centering\capstart
  \subfloat{\resizebox{\hsize}{!}{\includegraphics{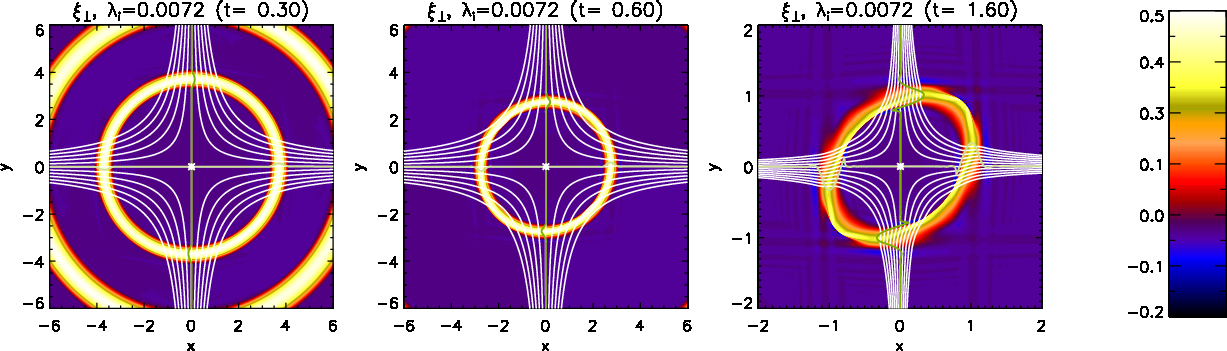}}}\\
  \subfloat{\resizebox{\hsize}{!}{\includegraphics{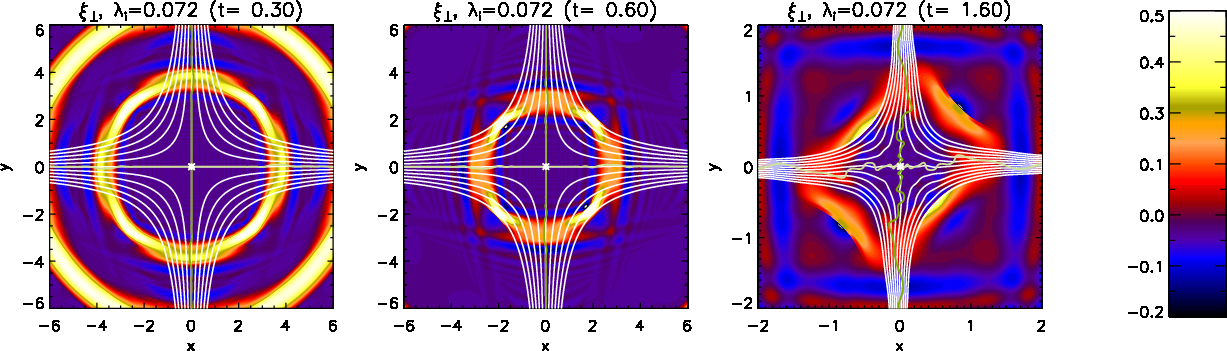}}}\\
  \subfloat{\resizebox{\hsize}{!}{\includegraphics{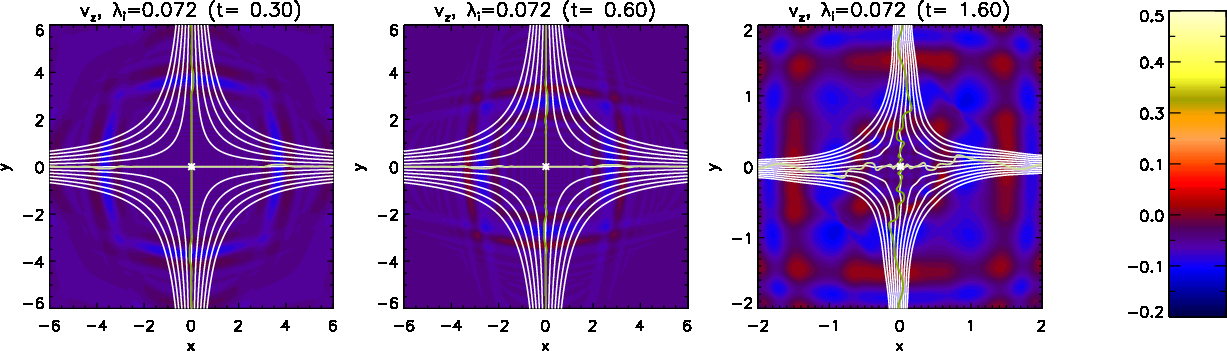}}}
\caption{Contours of $\xi_\perp$ (top and middle) and $v_z$ (bottom) seen through snapshots in time, for $\lambda_i=0.0072$ (top), and $\lambda_i=0.072$ (middle and bottom). Each case is overplotted with selected flux surfaces (contours of $A_z$ - in white), the lines of $B_x=0$ (dark green) and $B_y=0$ (light green), and also the location of the null-point (indicated by a white star). (NB. while the first two columns display $(x,y)\in[-6,6]$, the final column zooms in to illustrate a range of only $(x,y)\in[-2,2]$ of the entire simulated range, $(x,y)\in[-20,20]$).}
   \label{fig:annulusVperp}
\end{figure*}
 In particular, the behaviour of the pulse in the $\lambda_i=0.0072$ regime demonstrates many similarities with the equivalent MHD case; the inner pulse retains its initial profile, and is subject to refraction as it approaches the null point across the magnetic field lines. With no initial parallel velocity ($\xi_{||}=0$) and our choice of equilibrium field (Eq.~\ref{eq:EqConfig}), the system naturally develops an asymmetry in wave-speed \citep[due to the creation of a ``background inflow'' in the upper-left and bottom-right quadrants, while a ``background outflow'' is seen in the upper-right and bottom-left quadrants; see Fig.~4 of][where vertical and horizontal asymmetries are obtained due to the same effect]{paper:McLaughlinetal2009}. Thus, the initial velocity profile causes the peak of the pulse to propagate faster than the edges in the upper-left and lower-right quadrants, while the pulse edges travel faster than the peak in the upper-right and lower-left quadrants. These wave-speed asymmetries become discontinuous, and lead to the formation of fast-oblique (and perpendicular) shocks, as identified by \citet{paper:McLaughlinetal2009}. These shock fronts are associated with abrupt changes in density and temperature which coincide with discontinuities in the magnetic field, as seen in Fig.~\ref{fig:cutsf}.
\begin{figure*}[t]
  \centering\capstart
  \subfloat{\label{subfig:shockH1}\resizebox{\hsize}{!}{\includegraphics{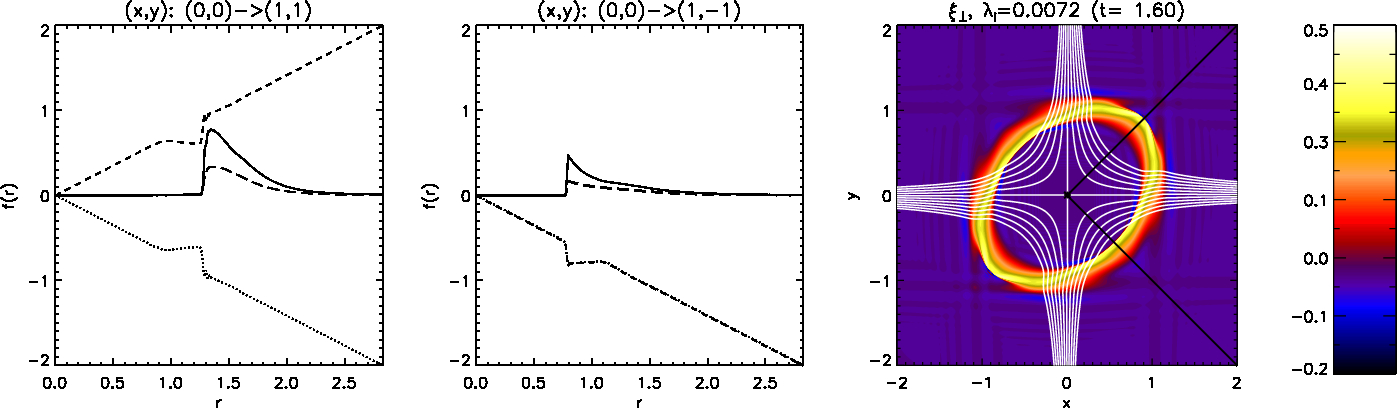}}}\\
  \subfloat{\label{subfig:shockHH1}\resizebox{\hsize}{!}{\includegraphics{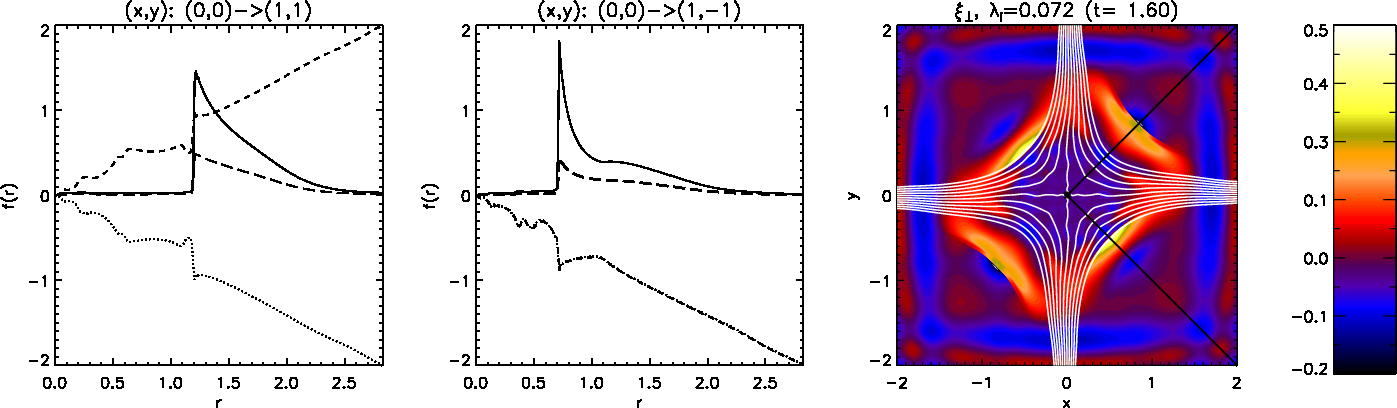}}}
\caption{Plots illustrating the jump conditions across the shocks in the system at $t=1.6\tau_A$ for different ion skin-depth values. Cuts through the shock wave front illustrate radial variations in $B_x$ (dotted line), $B_y$ (short dashes), gas pressure $p$ (solid line) and temperature $T$ (long dashes) for a radius $r=\sqrt{x^2+y^2}\in[0,2\sqrt{2}]$ ($p$ and $T$ are enhanced by a factor of $50$ to aid visual comparison). The positions of the cuts are indicated in the third column by the diagonal black lines, seen together with background contours of $\xi_\perp$ and selected contours of $A_z$ (in white).}
   \label{fig:cutsf}
\end{figure*}

In Hall MHD, differences begin to arise in comparison to the MHD benchmark, which increase in proportion to the value of $\lambda_i$. \citet{paper:Threlfalletal2011} demonstrated that the inclusion of the Hall term in the generalised Ohm's law causes an initially shear {\A} wave pulse (acting along a uniform magnetic field) to decouple into whistler and ion-cyclotron wave components, each having opposite circular polarisation. In this experiment, we see that the initially fast-mode wave pulse also becomes decoupled in this way, due to the inclusion of the Hall term. While much of the behaviour seen in the $\lambda_i=0.0072$ simulations matches that seen in the MHD benchmark case, the second and third rows of Fig.~\ref{fig:annulusVperp} show that increasing the effect of the Hall term also increases the amount of coupling to the shear velocity component, $v_z$. The asymmetry of wave-speed (described earlier) is also retained in simulations with $\lambda_i=0.072$, however the coupling to other wave modes and enhanced dispersion (which accompany the Hall term) limit the formation of the fast-oblique parts of the shock front. The second row of Fig.~\ref{fig:annulusVperp} illustrates this, showing that the shock front begins to take on the shape of the local magnetic field in this limit. The remaining wave annulus is concentrated at the centre of the four quadrants (where it remains exactly perpendicular to the local magnetic field, i.e. only the perpendicular shock fronts remain). Figure~\ref{fig:cutsf} shows that cuts through these shock fronts now reveal much larger density and temperature changes than in the $\lambda_i=0.0072$ (and MHD benchmark) case(s).

Note that the diffraction-like pattern recovered in Figs.~\ref{fig:annulusVperp} and~\ref{fig:cutsf} are not related to any contact with a boundary, as these figures represent only a small region of the entire simulated domain, $(x,y)\in[-20,20]$.

\section{X-point collapse and evolution of magnetic topology}\label{sec:Nulls}
The response of the magnetic field to the inward-travelling pulse and the evolution and formation of current sheets greatly depends on the normalised ion-skin depth, $\lambda_i$. In the case where $\lambda_i=0.0072$ (as with the MHD benchmark) the evolution of the system largely follows that outlined by \citet{paper:McLaughlinetal2009}; fast magnetic shock waves, formed in opposite quadrants, travel in towards the null, significantly deforming the magnetic field-lines as they pass. The initial interaction between the wave annulus and the null may be seen in Fig.~\ref{fig:Hcusp}.

\begin{figure*}[t]
  \centering\capstart
  \subfloat{\resizebox{\hsize}{!}{\includegraphics{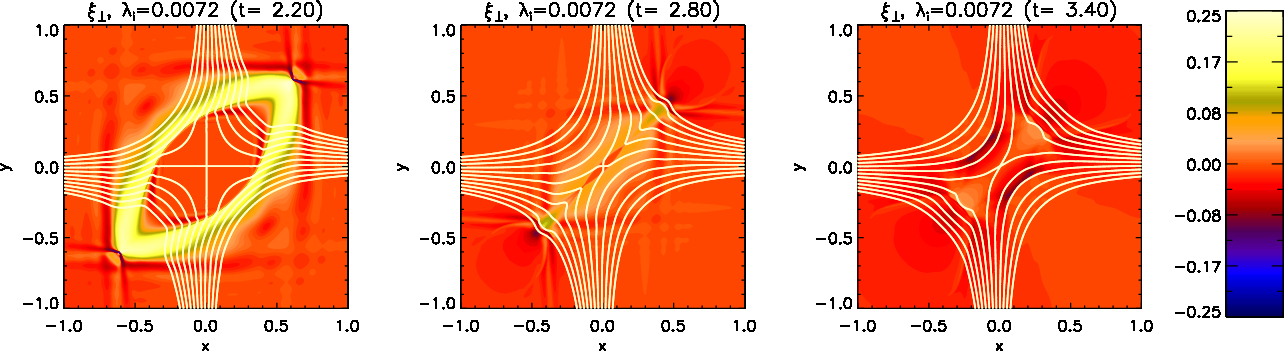}}}
\caption{Illustration of pulse evolution and X-point collapse in simulations where $\lambda_i=0.0072$, following the formation of magnetic shocks; contours of $\xi_\perp$ are seen at various times and overlaid with selected contours of $A_z$ (in white).} 
   \label{fig:Hcusp}
\end{figure*}
In this limit (and in MHD), the outer edges of the shock front begin to overlap at $t\sim2\tau_A$ [at approximately $(x,y)=(1,1)$ and $(x,y)=(-1,-1)$], leading to the formation of hot ``cusp'' jets which heat the plasma (in the opposite quadrants to those where the original fast shocks predominantly formed). The centre of the fast shock-fronts reach the null and begin to pass through each other at $t\approx2.9\tau_A$, whereupon the local magnetic topology has deformed into a thin current sheet-like structure. The fact that the wave is able to reach and pass through the null is entirely due to our choice of a large initial wave amplitude and the nonlinear effects which result; a small amplitude wave would slow while approaching the null but never actually reach it (as $\ca\rightarrow0$).

After $t\approx2.9\tau_A$, the original fast shock-waves begin to pass through the null, allowing wave energy to travel out from the centre. This energy is removed from the system before any waves reach the simulation boundaries, due to the presence of the damping layer. Meanwhile the hot cusp-jets continue to grow and spread out, leading to the formation of further magnetic shocks which, in combination with a local increase in gas pressure, act to compress the initial current sheet and wedge open the separatrices. This ultimately causes the magnetic field to overshoot its equilibrium configuration, leading to the formation of a second current sheet, of opposite orientation in $J_z$ and perpendicular to the original current sheet (i.e. from top-left to bottom right). The system then relaxes through a series of these current sheets, a process which is described in detail in Sect.~\ref{sec:Oscill}.

Increasing $\lambda_i$ by a factor of $10$ highlights several marked differences with evolution outlined above, both in the way in which the wave annulus interacts with the local field and the field response. As mentioned earlier, the fast wave is not only coupled to a shear component of velocity through the inclusion of the Hall term, but also now contains both whistler and ion-cyclotron waves. In particular, dispersive whistler waves allow information to travel faster than both the local {\A} and fast MA wave speeds. In the $\lambda_i=0.072$ simulations, whistler waves are responsible for rapid oscillations of the magnetic field near the origin, at a much earlier time than the $\lambda_i=0.0072$ results (and also in MHD); these early oscillations are demonstrated in the contours of $B_x=0$ and $B_y=0$ in the third column of Fig.~\ref{fig:annulusVperp}. As the oscillations grow in amplitude, multiple additional nulls are created in a small region close to the origin (of radius $r\sim0.5$). The additional nulls appear in pairs (containing both an X-type and O-type null); each new null pair is evidence of local magnetic reconnection. Every null point is located to sub-grid resolution accuracy using a linear interpolation algorithm \citep[described in detail in][]{paper:HaynesParnell2007}, with the results catalogued in Table~\ref{tab:HHnulltab} according to phases during the simulations which display similar properties (such as number of nulls, local current sheet orientation, etc.). As we are using interpolation, the precise location of the nulls must be regarded as an approximation (with the error on the order of a fraction of the grid resolution).

\begin{table}[t]
	\caption{Configuration of Nulls}
     \label{tab:HHnulltab}
     \centering
          \begin{tabular}{lclllcc}\hline\hline
	Phase	& Nulls&X's&O's& Central & Time & C-S \\
	&(X+O)&&&Null&&Orientation \\
              \hline
1	& 1&1&0	& X	& $<$2.3	& n/a \\
2	& 5&3&2	& X	& 2.3		& $/$ \\
3	& 7&4&3	& O	& 2.4		& $\backslash/\backslash$\\
4	& 5&3&2	& X	& 2.5		& $\backslash$\\
5a	& 5&3&2	& X	& 2.60		& $\backslash$\\
5b	& 1&1&0	& X	& 2.61-2.64	& $\backslash$\\
5c	& 1&1&0	& X	& 2.65-2.66	& $/$\\
5d	& 3&2&1	& X$^\dagger$& 2.67	& $/$\\
5e	& 5&3&2	& X	& 2.68		& $/$\\
6	& 3&2&1	& O	& 2.69-3.4	& $/$\\
7	& 7&4&3	& O	& 3.5-3.6	& $/$\\
8	& 5&3&2	& O$^\dagger$& 3.7	& $/$\\
9	& 7&4&3	& O	& 3.8		& $/$\\
10	& 3&2&1	& O	& 3.9-4.5	& $/$\\
11	& 1&1&0	& X	& $>$4.6	& $\ast$\\
              \hline
          \end{tabular}
\tablefoot{Summary of the configuration of null-points in the $\lambda_i=0.072$ simulations, broken down into phases which display similar properties. In each phase, the number of nulls in the system and their type are given, along with the central null type, the time over which the phase occurs and the orientation of any local current sheets ($/$ infers a current sheet oriented along the line $y=x$, and $\backslash$ along the line $y=-x$). [NB. $\dagger$ denotes a null asymmetry - the ``central null'' refers to the null closest to the origin in these instances; $\ast$ denotes that the current sheet orientation alternates periodically via oscillatory reconnection].}
\end{table}
Many of the early phases containing multiple nulls (phases~2-4 of Table~\ref{tab:HHnulltab}) are highly dynamic; the number and position of nulls varies rapidly, and phases (with the same number of nulls) are relatively short-lived. All null points which arise as a result of the dispersive whistler component exist only on or close to the lines $y=x$ or $y=-x$, while any current sheets which form in the system also evolve rapidly and are short-lived. The formation of the whistler-generated nulls may be seen in the first row of Fig.~\ref{fig:HHnulls}, with subsequent rows illustrating the later evolution of nulls and currents in the system.

It is interesting to note that the behaviour of the MHD/$\lambda_i=0.0072$ simulations is encapsulated by only 3 of the phases outlined in Table~\ref{tab:HHnulltab}. In these cases, the initial configuration (phase~1) is retained until the arrival of the fast-shocks at the null (phase~5c), after which the system relaxes through the oscillatory reconnection phase (11). The additional phases which occur only in the $\lambda_i=0.072$ case are generated by the whistler components (for phases~2-5c), and following the arrival of the pulse, by a combination of whistler and ion cyclotron (fast and shear) wave components (phases~5d-10).

\begin{figure*}[p]
  \centering\capstart
  \subfloat{\label{subfig:HHnullsa}\resizebox{\hsize}{!}{\includegraphics{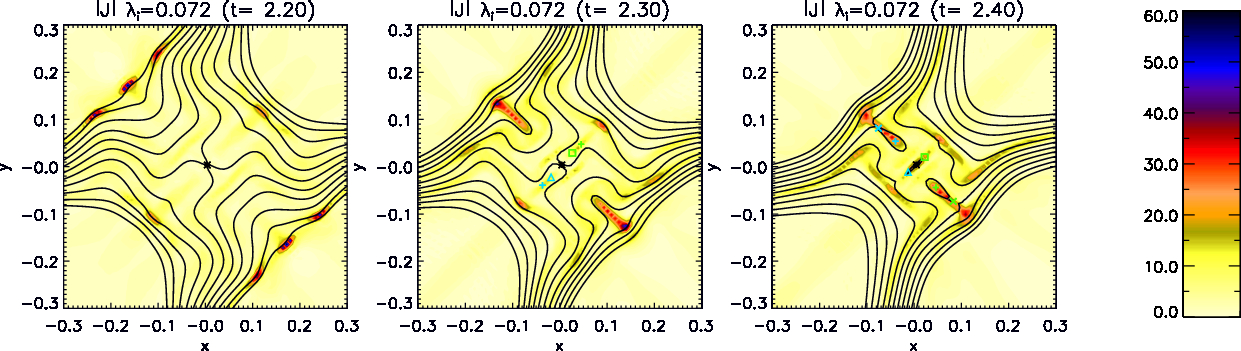}}}\\
  \subfloat{\label{subfig:HHnullsb}\resizebox{\hsize}{!}{\includegraphics{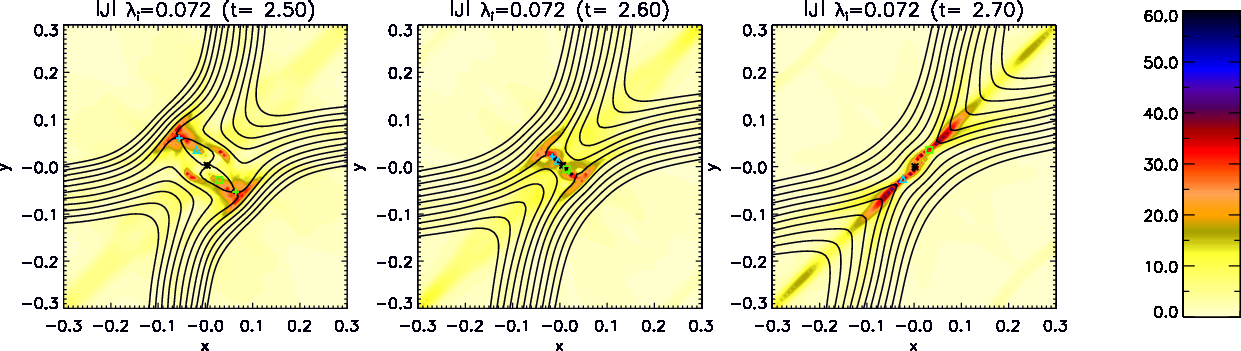}}}\\
  \subfloat{\label{subfig:HHnullsc}\resizebox{\hsize}{!}{\includegraphics{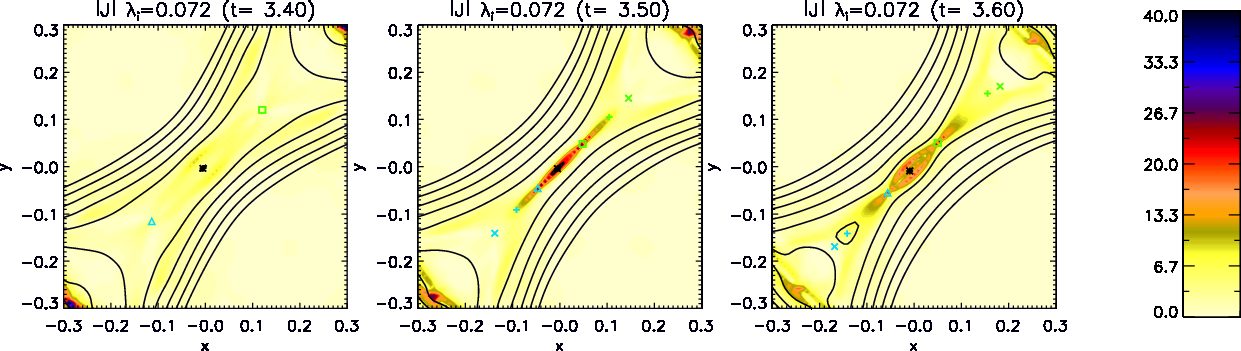}}}\\
  \subfloat{\label{subfig:HHnullsd}\resizebox{\hsize}{!}{\includegraphics{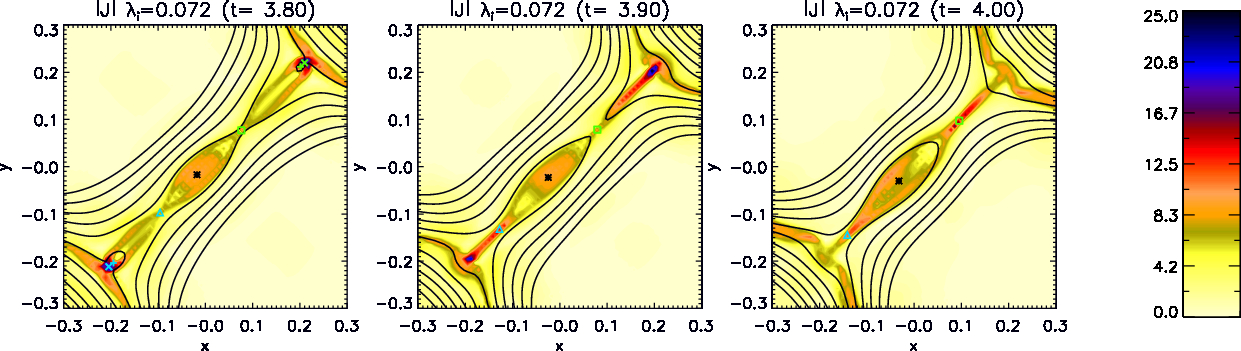}}}
\caption{Contours of current, $|{\bf{J}}|$, illustrating snapshots of the $\lambda_i=0.072$ simulations, focussing closely on behaviour near the null, highlighting the presence and locations of multiple nulls, overlaid with selected contours of $A_z$ (in black). Each null is assigned a symbol and a colour based on the null position relative to the central null, which is indicated by a black star. Nulls above this along the contour of $B_y=0$ would be seen in green and those below are shown in light blue. The distance along the $B_y=0$ line is indicated by the symbol - a snapshot containing 7 nulls would thus display, from left to right, $ {\color{Cyan}{\times + \triangle}} \ast {\color{green}{\square + \times}}$.} 
   \label{fig:HHnulls}
\end{figure*}
Using Table~\ref{tab:HHnulltab} and Fig.~\ref{fig:HHnulls}, the remaining evolution of the $\lambda_i=0.072$ system may be described as follows; the arrival of two perpendicular magnetic shocks at the origin (originating from the upper-left and lower-right quadrants) halts the rapid oscillation of the field in the local vicinity. This is illustrated in phase 5b of Table~\ref{tab:HHnulltab}, highlighting that the shock fronts cause all additional null pairs to recombine, leaving a single X-type null at the origin. The arrival of these shock fronts at the remaining null, between $2.64\tau_A<t<2.65\tau_A$, reverses the orientation of the currents generated by earlier whistler oscillations (seen in the second row of Fig.~\ref{fig:HHnulls}). The remaining shock waves then continue to pass through the origin and each other, forming a relatively stable current sheet oriented along the line $y=x$ (as with the previous MHD/$\lambda_i=0.0072$ simulations). Unlike the previous cases, multiple null pairs are then created, at a range of locations which all lie along the current sheet, up to a maximum of 6 additional nulls at any one time. At $t\approx2.69\tau_A$ (phase~6 of Table~\ref{tab:HHnulltab}), the appearance of a null pair close to the origin causes the existing X-type null to be displaced from the origin, with a new O-type null-point taking its place at the centre. The central null remains as O-type until $t\approx4.5\tau_A$. Islands of current form around the O-type nulls in the system, and, in particular, at this long-lived central O-type null (demonstrated by the final row of Fig.~\ref{fig:HHnulls}).
 
Phase 11 of Table~\ref{tab:HHnulltab} marks destruction of the additional null pairs, leaving a single X-type null at the origin. This coincides with the arrival of the second (slower) pair of shock waves from the upper-right and lower-left quadrants (the formation of which may be seen in the second rows of Figs.~\ref{fig:annulusVperp} and~\ref{fig:cutsf}). These shocks arrive much later than the initial shocks which cause the initial X-point collapse (due to the wave-speed asymmetry discussed earlier), and cause any remaining additional nulls to merge. At this point the system returns to the original configuration; a single X-type null-point at the origin about which forms a series of oppositely oriented current sheets, as with previous MHD/$\lambda_i=0.0072$ cases. 

\section{Oscillatory relaxation}\label{sec:Oscill}
Both in the MHD and Hall MHD simulations, the majority of the initial transient wave features have left the system by $t\approx10\tau_A$, and no longer disturb the null. At this point, the perturbed field begins to evolve back to a force-balanced state through a series of oppositely oriented current sheets, reconnecting magnetic flux at the single X-type null-point as it relaxes \citep[see e.g.][]{paper:CraigMcClymont1991,paper:McLaughlinetal2009}.

\begin{figure}[t]
 \centering \capstart
\resizebox{\hsize}{!}{\includegraphics{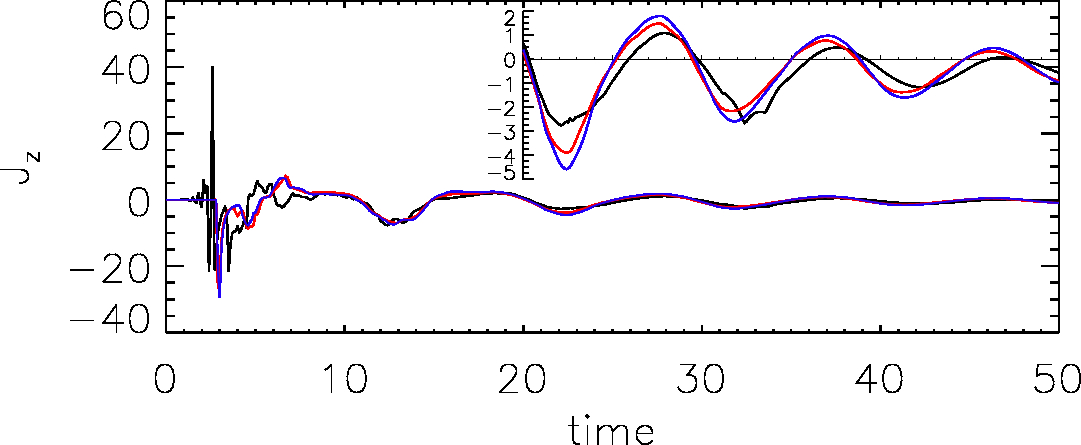}}
\caption{Comparison of the time evolution of the out-of-plane current, $J_z$, at the central null. The black line indicates the value of $J_z$ for the central null in the $\lambda_i=0.072$ simulations, overlaid with the $\lambda_i=0.0072$ results (in red) and the MHD results (in blue). The insert displays the behaviour of all 3 cases at later times but over a much smaller range of $J_z$ in the y-axis).} 
\label{fig:cnullevol}
\end{figure}
In the MHD/$\lambda_i=0.0072$ investigations, only a single null is ever present, located at the origin, where no initial current density is recorded prior to the arrival of the fast wave (at approximately $\sim2.9\tau_A$). \citet{paper:Threlfalletal2011} showed that the dispersion relation derived for the long wavelength Hall MHD regime yields a modified form of the MHD dispersion relation. In this limit, the whistler and {\A} speeds are well matched, but moving to the short wavelength Hall MHD limit rapidly increases the whistler speed in comparison to the {\A} speed. Thus the MHD/$\lambda_i=0.0072$ simulations begin to destabilise the magnetic field at the origin at approximately the same time (as seen in Fig.~\ref{fig:cnullevol}) but this destabilisation occurs much earlier in simulations with $\lambda_i=0.072$.

The arrival of the fast-wave at the null (seen in Fig.~\ref{fig:Hcusp} for the case where $\lambda_i=0.0072$) forms a strong current sheet at the origin. This is seen as the deep initial minima in the blue and red curves of Fig.~\ref{fig:cnullevol}. This current sheet is then destabilised both by the remaining fast-wave exiting the system, and the ``wedging open'' of the magnetic field by the cusp jets described earlier \citep[and studied in detail by][]{paper:McLaughlinetal2009}. The preferred direction for the generation of a current sheet \citep[noted by][]{paper:McLaughlinetal2009} results from the orientation of the initial current sheet (formed by the arrival of the fast shocks). This orientation also determines in which quadrants plasma pressure is increased; the fast/perpendicular shocks compress and heat plasma, particularly to the left and right of the current sheet. The plasma pressure enhancement which results then exerts a force to open the compressed X-point quadrants, and return the system to equilibrium. The first peak of the MHD/$\lambda_i=0.0072$ curves in Fig.~\ref{fig:cnullevol} signals the formation of a second current sheet, of opposite orientation, generated by the field overshooting its initial equilibrium configuration. While minor cusp jets are also observed during this phase (once again causing the magnetic field to widen), they are seen to be much weaker and are not observed in any further oscillation cycles; hence they do not play an essential role in the subsequent oscillatory process. 

\begin{figure*}[t]
 \centering \capstart
 \resizebox{\hsize}{!}{\includegraphics{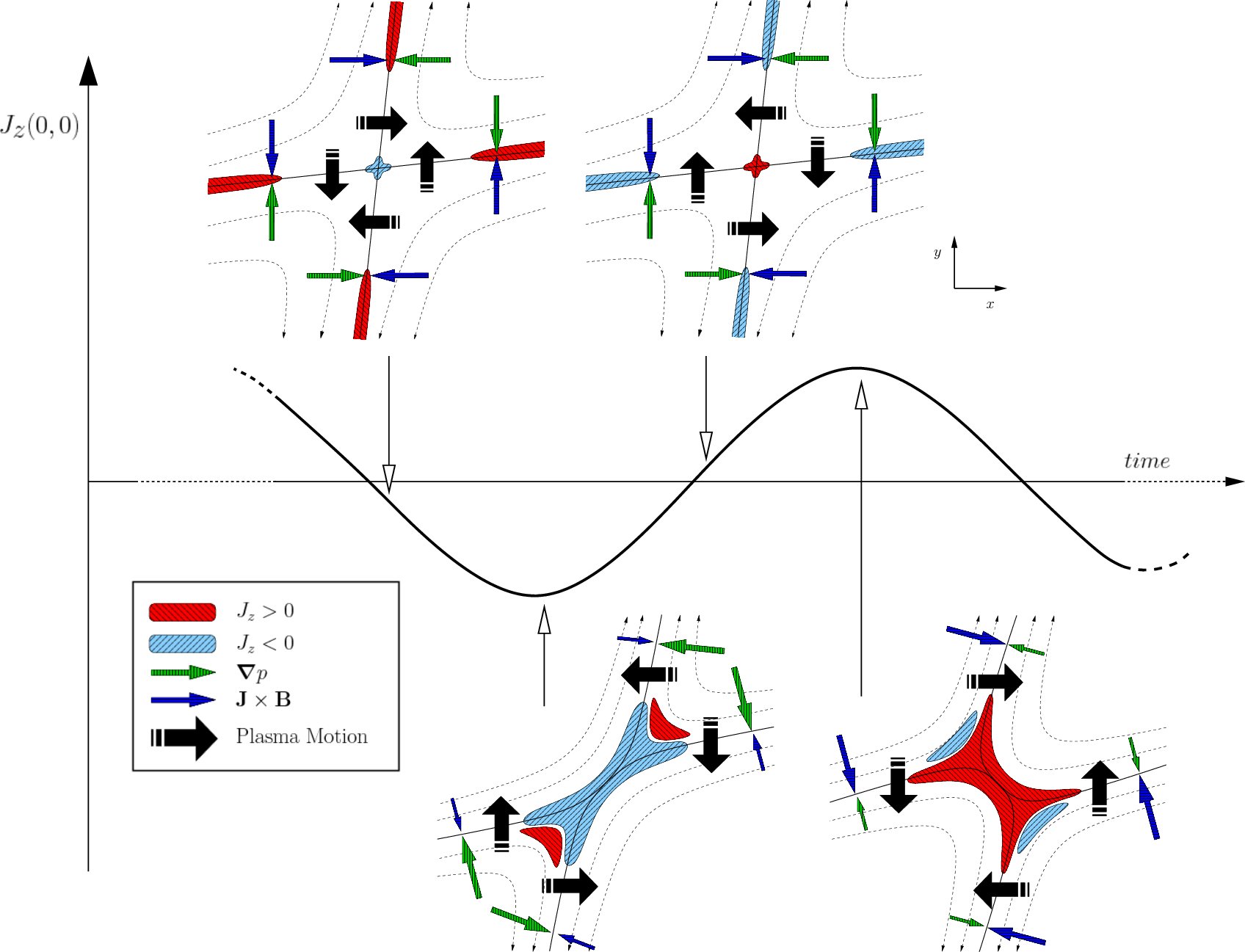}}
 \caption{Cartoon illustrating the cycle of forces generated during the oscillatory reconnection phase (for key to symbols, see legend).} 
 \label{fig:cartoon}
\end{figure*}
The oscillatory nature of the experiment (at stages where the cusp jets are no longer seen) is generated by a competition of Lorentz and gas pressure forces, each in turn restoring an overshoot of the equilibrium configuration brought on by the other. A simplified description of this competition of forces may be seen in Fig.~\ref{fig:cartoon}. Of the Lorentz force components, magnetic pressure dominates in these experiments (magnetic tension aids the compression of magnetic field lines in the compressed quadrants of the X-point as the current sheet forms). As each successive overshoot is smaller than the last, the system is ultimately able to relax.

\begin{figure*}[t]
 \centering \capstart
 \resizebox{\hsize}{!}{\includegraphics{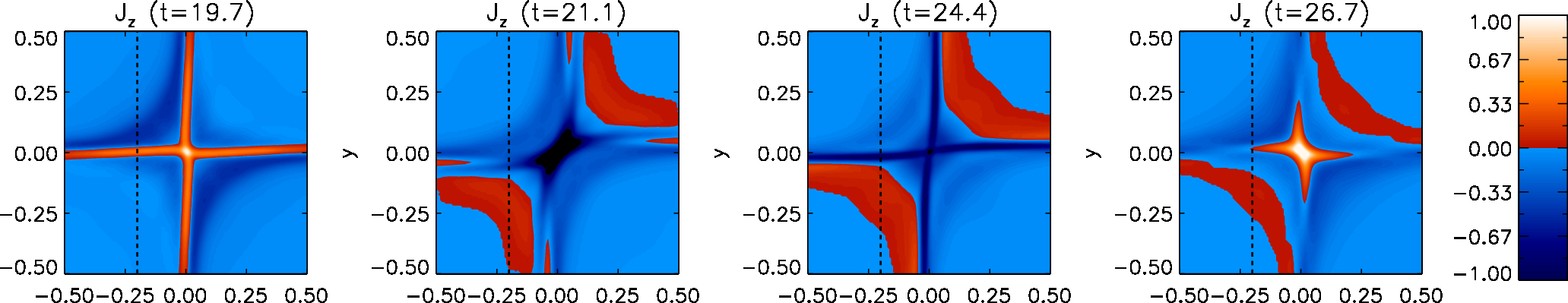}}
 \caption{Illustration of current sheets generated during oscillatory phase. Each snapshot displays normalised perpendicular current, $J_z$, in a region near the X-point, $x,y\in[-0.5,0.5]$, at selected times during the oscillatory cycle (chosen for comparison with Fig.~\ref{fig:fvcombo}, which displays quantities along a cut indicated in each frame by a dashed black line).} 
 \label{fig:jzcombo}
\end{figure*}
\begin{figure}[t]
 \centering \capstart
 \capstart\subfloat{\label{subfig:fcombo}\resizebox{\hsize}{!}{\includegraphics{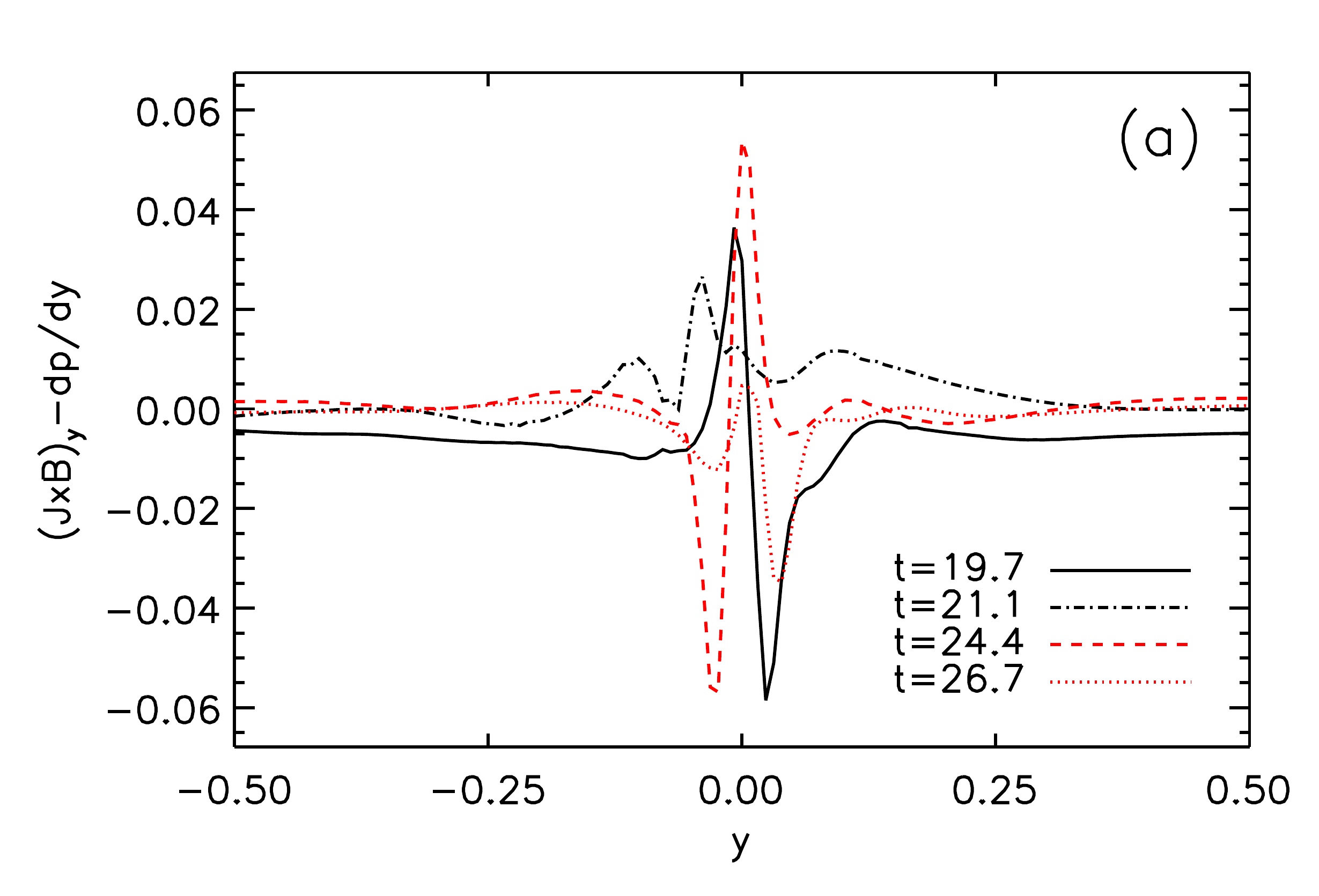}}}\vspace{-5mm}\\
 \capstart\subfloat{\label{subfig:vcombo}\resizebox{\hsize}{!}{\includegraphics{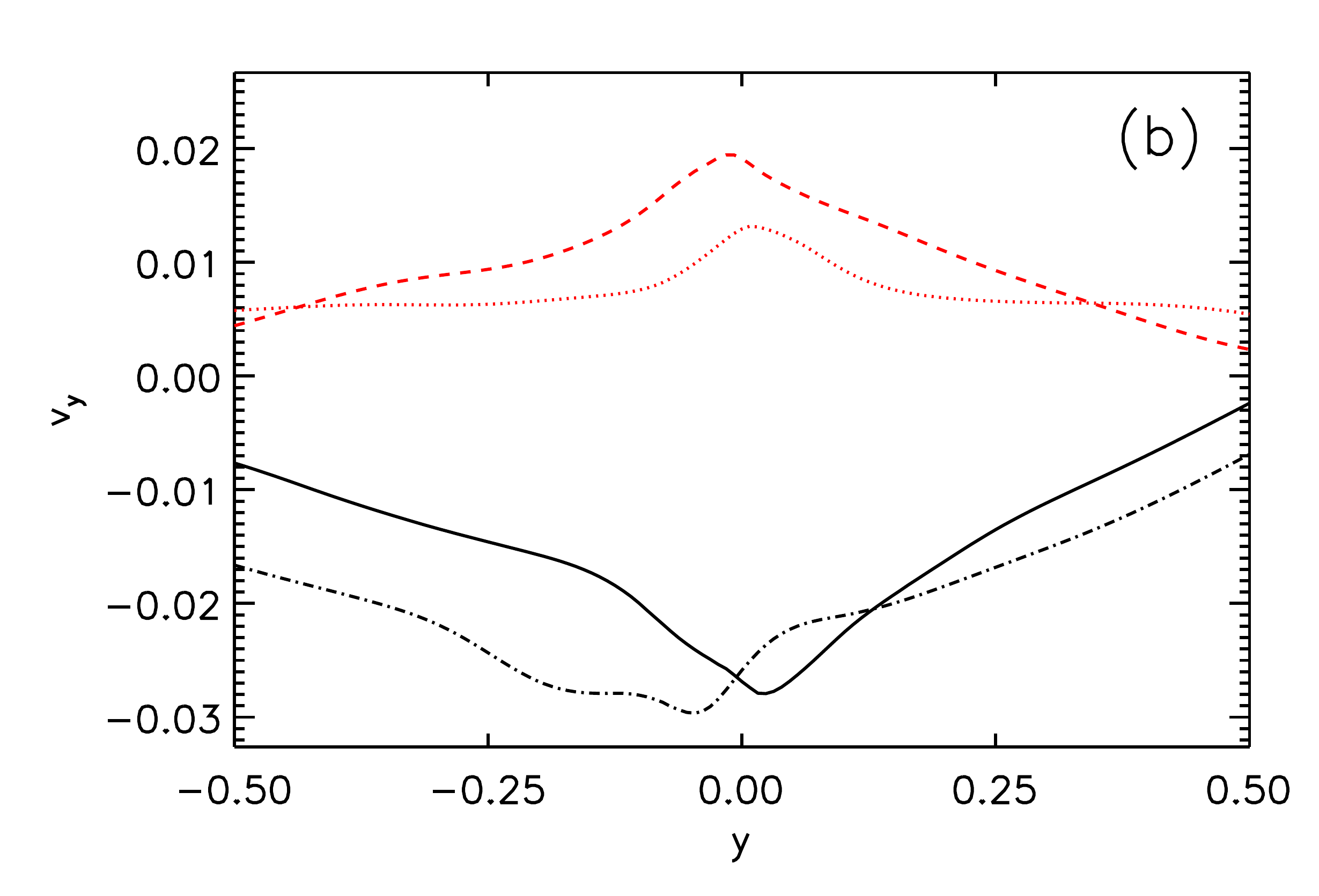}}}
 \caption{$y$-components of \protect\subref{subfig:fcombo} a summation of Lorentz and gas pressure forces and \protect\subref{subfig:vcombo} velocity in a cut along the $y$-axis at selected times; the timing of each cut is chosen for comparison with the local current configurations demonstrated in Fig.~\ref{fig:jzcombo} (where the location of each cut is also illustrated by a dashed black line). One half of the oscillatory cycle is indicated in red, with the remaining half in black (at times indicated in the legend).} 
 \label{fig:fvcombo}
\end{figure}
We also present supporting evidence for this competition of forces, in the form of snapshots of the local current configuration near the null (Fig.~\ref{fig:jzcombo}) and a summary of the forces and velocity flow in a cut through one of the separatrices (Figs.~\ref{subfig:fcombo} and \ref{subfig:vcombo} respectively), sampled at the same time. Beginning at $t=19.7\tau_A$, the chosen cycle begins with Lorentz and gas pressure forces approximately balanced and a strongly negative plasma velocity (Fig.~\ref{fig:fvcombo} - solid lines), while Fig.~\ref{fig:jzcombo} shows that the plasma is close to its original configuration (and that a previous strong positive $J_z$ current sheet at the origin is diminishing as the field returns towards equilibrium). A large plasma velocity (acting downward along the chosen cut) generates inertia, which drags field lines past their equilibrium configuration, compressing the plasma in the lower-left and upper-right quadrants. Subsequently a gas pressure response is generated in these quadrants. Along the cut, a strong gas pressure force acting upwards dominates the competition of forces at $t=21.1\tau_A$ in Fig.~\ref{subfig:fcombo} (dot-dashed lines). Concurrently, a current sheet (of negative $J_z$ and of opposite orientation to the first) has formed in the corresponding panel of Fig.~\ref{fig:jzcombo}. The force imbalance slows the plasma; in Fig.~\ref{subfig:vcombo}, the plasma velocity reaches its largest negative value and reverses sign as it accelerates back towards an equilibrium state. Having once more returned to a near-equilibrium field configuration (at $t=24.4\tau_A$), Fig.~\ref{subfig:fcombo} (dashed lines) shows Lorentz and gas pressure forces are near-balanced, but the large upward velocity in~\ref{subfig:vcombo} now carries plasma past equilibrium in the opposite direction (compressing the upper-left and lower-right quadrants). Note that the velocity seen here is approximately at maximum for this particular phase of the cycle, but is much smaller than the equivalent maximum downward velocity generated earlier; each successive peak is smaller than the last as each overshoot gradually reduces in size as the system relaxes. In the final snapshot (at $t=26.7\tau_A$), Fig.~\ref{fig:jzcombo} shows the formation of a positive $J_z$ current sheet near the origin, while the plasma velocity has begun to reduce (Fig.~\ref{subfig:vcombo}, dotted lines). The reduction in plasma velocity is caused by a second force imbalance (where forces along the cut are now dominated by contributions from the Lorentz force, acting downward in~\ref{subfig:fcombo}). Plasma velocity continues to reduce and the forces return to balance once more; with the plasma returning towards its equilibrium configuration the cycle repeats.

\section{Nature of reconnection}\label{sec:Reco}
The reconnection which takes place in our system may be broadly grouped into two categories, associated with either the initial X-point collapse or the later oscillatory relaxation. While the oscillatory reconnection process is very similar in all 3 experiments, the initial X-point collapse differs greatly as the influence of the Hall term increases. 

For the $\lambda_i=0.0072$/MHD benchmark simulations, the wave annulus reaches the null at approximately $t\sim2.8\tau_A$, causing the initial departure from zero of $J_z$ at the null, plotted in Fig.~\ref{fig:cnullevol} (red and blue lines), and thus marking the onset of reconnection. Following this, the reconnection associated with the initial X-point collapse follows the description given in \citet{paper:McLaughlinetal2009}. The initial current sheet (aligned with the line $y=x$) is ``wedged open'' by the formation of cusp jets, forming a second current sheet perpendicular to the first. This second current sheet (aligned with the line $y=-x$) is maintained until the oscillatory reconnection phase (discussed earlier and below) starts at around $t\sim10\tau_A$. In simulations with $\lambda_i=0.072$, the onset of reconnection occurs much earlier, when a rapid oscillation of the field at the origin (due to the whistler wave component) creates multiple X-type and O-type null-points, which by their very existence provide evidence of local magnetic reconnection. In Fig.~\ref{subfig:scattera}, we show how the number of nulls and the local current at each null varies in both Hall MHD experiments. The arrival of perpendicular magnetic shocks causes the multiple nulls to recombine, leaving a single X-type null about which a single large current sheet forms. Subsequently, further nulls are generated within this current sheet, which is ultimately collapsed by a second pair of shocks; these shocks cause the nulls to recombine once more, after which only a single null remains at the centre, with the oscillatory phase once more setting in at around $t\sim10\tau_A$.

\begin{figure}[t]
 \centering
 \capstart\subfloat{\label{subfig:scattera}\resizebox{\hsize}{!}{\includegraphics{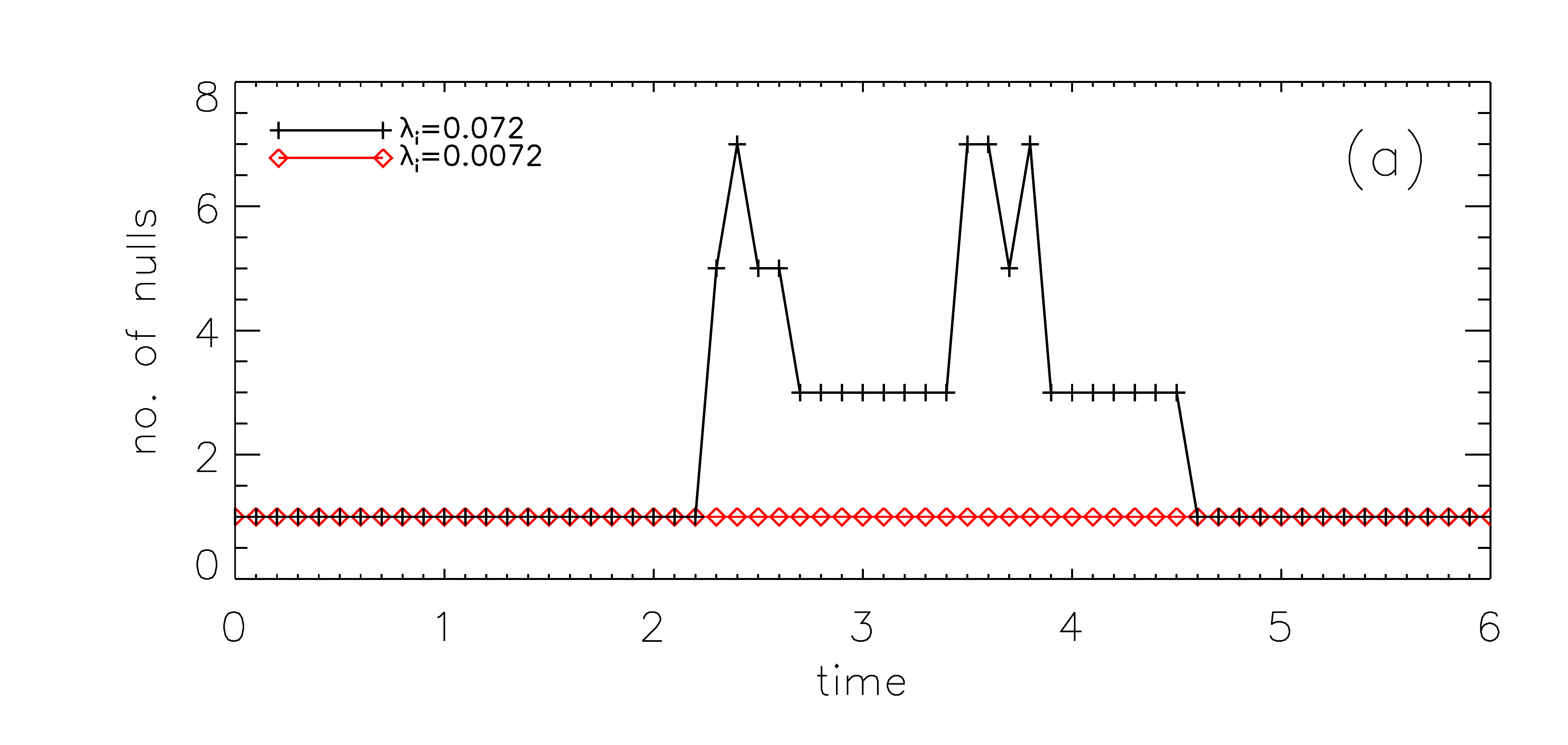}}}\vspace{-5mm}\\
 \capstart\subfloat{\label{subfig:scatterb}\resizebox{\hsize}{!}{\includegraphics{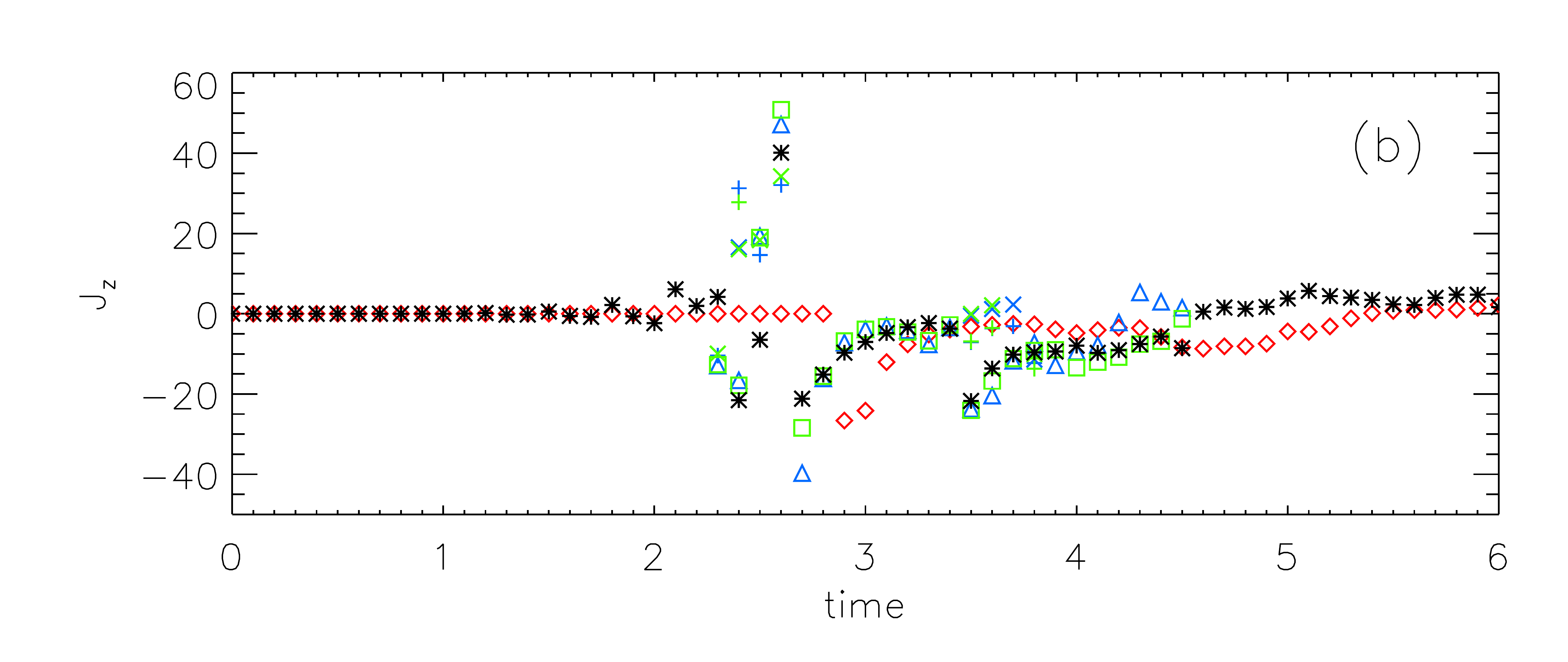}}}\vspace{-5mm}\\
 \capstart\subfloat{\label{subfig:scatterc}\resizebox{\hsize}{!}{\includegraphics{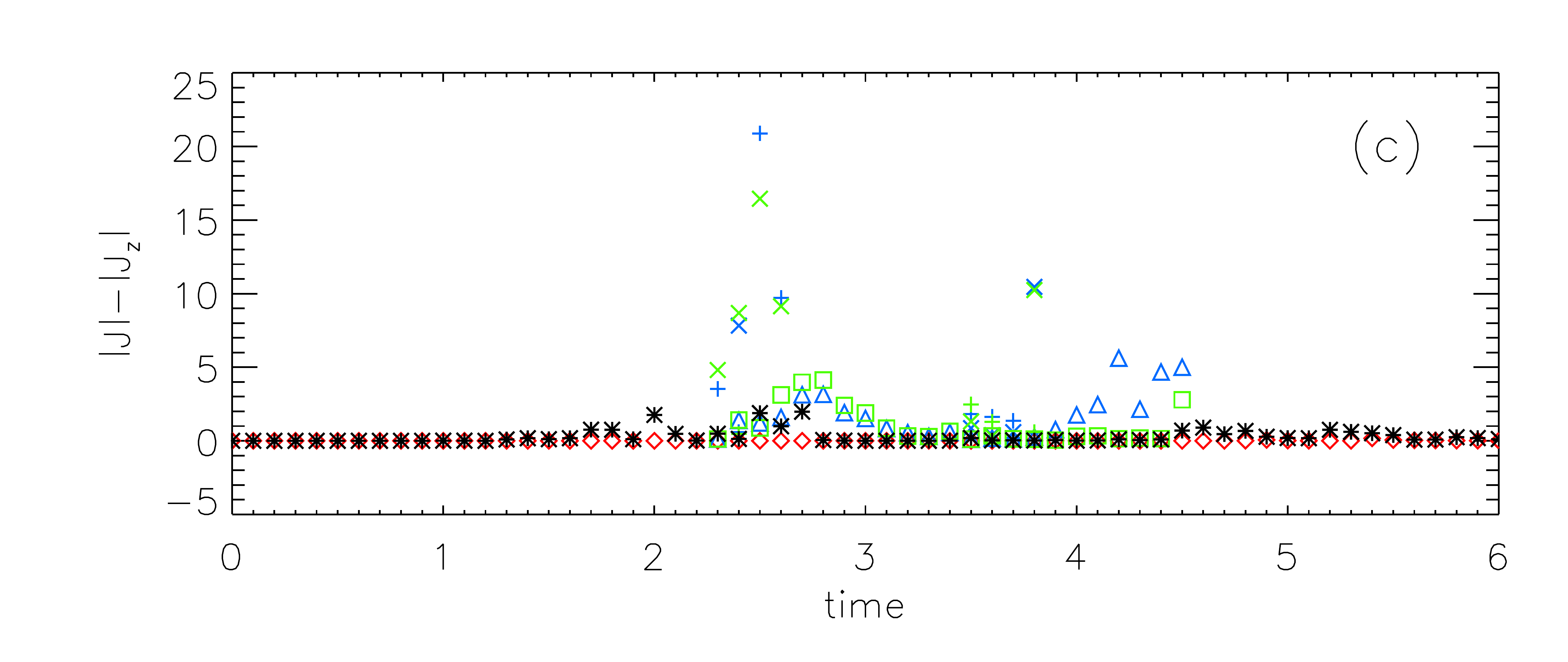}}}
\caption{Figure~\protect\subref{subfig:scattera} indicates the number of nulls present in the Hall MHD simulations. The remaining figures illustrate the evolution of current at each null; the values of \protect\subref{subfig:scatterb} $J_z$ and \protect\subref{subfig:scatterc} $|{\bf{J}}|-|J_z|$ at an individual null are shown, comparing the single null case ($\lambda_i=0.0072$, in red) with values at every null in the $\lambda_i=0.072$ simulations. When multiple nulls are present, each null is uniquely identified by a combination of colour and symbol, identical to that outlined in Fig.~\ref{fig:HHnulls} [e.g. 7 nulls, ordered according to location along the $B_y=0$ line, would be identified by $ {\color{Cyan}{\times +  \triangle}} \ast {\color{green}{\square + \times}}$].}
\label{fig:nullstats}
\end{figure}
During the initial X-point collapse phase, large currents (both perpendicular and planar) are often associated with many of the nulls in the system (regardless of type) as indicated by Figs.~\ref{subfig:scatterb} and~\ref{subfig:scatterc}. These currents provide insight into the local reconnection rate at each null. It is clear from~\ref{subfig:scatterb} and~\ref{subfig:scatterc} that the currents at the nulls during the $\lambda_i=0.072$ case are typically larger that those during the $\lambda_i=0.0072$/MHD cases, indicating that the rate of reconnection increases with the value of $\lambda_i$, as one might expect.

How each local rate contributes to a global rate (for a system containing multiple X-type and O-type nulls which continually change in number and position) is unknown. In 2D, reconnection can only occur at X-points, while annihilation/dissipation and generation of magnetic flux is associated only with O-points \citep[pp.~41,][]{book:BirnPriest}. Annihilation and generation of magnetic flux are distinct from reconnection since they involve a loss or a gain of magnetic flux rather than a transfer of flux between topologically distinct regions. In the $\lambda_i=0.072$ simulations, where there are multiple nulls, reconnected flux from one X-point may reconnect again at another X-point, repeating this process several times. It is still an open question as to whether repeated reconnection should or should not be taken into account when determining an overall (i.e. global) rate of reconnection. Additionally, flux pile-up may occur at O-points, which again may or may not affect a global reconnection rate. 

Calculation of a global reconnection rate through a summation of rates at individual X-points (for example) would be insufficient to describe all the reconnection which takes place. The sign of the reconnection rate (seen through the orientation of the local electric field) at each X-point varies significantly (according to Fig.~\ref{fig:nullstats}); nulls with similar rates which are opposed in sign would cancel if all rates were simply summed, obscuring the true amount of reconnection which has taken place. The variation of the position of each null also calls into question whether a single (global) diffusion region or several smaller diffusion regions are present in Hall MHD reconnective regimes. Finally, in some cases, a significant fraction of the total current at a null may be located in one or both of the planar components, $J_x$ and $J_y$, as seen in Fig.~\ref{subfig:scatterc}. How these currents affect reconnection in a 2D model requires further consideration.

The second category of reconnection, as already mentioned, is oscillatory and involves just a single null in all experiments considered. Despite the different skin depths in the experiments, both MHD and Hall MHD simulations yield nearly identical frequencies of current oscillation at the central null (albeit with a phase-difference in the $\lambda_i=0.072$ case, as seen in Fig.~\ref{fig:cnullevol}).

\citet{paper:CraigMcClymont1991} studied the oscillatory relaxation of a 2D X-point field following a perturbation with well defined azimuthal modes. They obtained analytical solutions which included an expression for the oscillation period of the lowest frequency, associated with the most slowly decaying eigenfunction of their system which relaxes through oscillatory reconnection. This analytical treatment, when performed for a purely azimuthal (i.e. $m=0$), fundamental radial mode ($n=0$), yields a period of oscillation for topological reconnection, which depends only on the Lundquist number:
\[
 \tau_{\rm osc}^{n=0}=2\ln{S}, \nonumber 
\]
where $\tau_{\rm osc}^{n=0}$ is in units of {\A} times, $\tau_A=l_0/\ca$. Substituting the Lundquist number for our experiment ($S=2000$) yields an oscillation period $\tau_{\rm osc}^{n=0}\approx15.2$, which is significantly higher than the period of oscillation seen from the oscillation of current at the null in Fig.~\ref{fig:cnullevol}, which we estimate to be $\tau_{\rm osc}\approx10$.

The smaller oscillation time in our simulations compared to the value given by the above equation may be explained by the fact that the latter applies only to the lowest frequency radial eigenmode of the system considered by \citet{paper:CraigMcClymont1991}. In contrast, the fast wave modelled in our simulation contains many radial eigenmodes, with a range of frequencies higher than that of the fundamental mode. Furthermore we employ damping layers and a non-circular boundary that differs from that of the system considered by \citeauthor{paper:CraigMcClymont1991}. Finally, nonlinear effects in our simulations cause the annulus to become azimuthally asymmetric, introducing coupling to finite $m$ modes, for which \citeauthor{paper:CraigMcClymont1991} did not compute any eigenfunctions.
It is therefore to be expected that while the overall oscillation period we obtain is of the same order as the fundamental mode period found by \citet{paper:CraigMcClymont1991}, the presence of higher frequency radial eigenmodes, nonlinear effects and the choice of boundary conditions means that the period in the simulation differs from that obtained by these authors.

\section{Discussion and conclusions}\label{sec:Conc}
Simulations of an X-type null point disturbed from equilibrium by an incident nonlinear fast wave annulus behave largely as described by \citet{paper:McLaughlinetal2009}, both in MHD and in Hall MHD simulations with a small ion skin depth value. Increasing the ion-skin depth by a factor of 10 shows significant differences, both in the evolution of the initial annulus and the response of the local magnetic environment. The initial fast wave annulus rapidly deforms, coupling to a shear component which may only travel along magnetic field lines. The oblique parts of the shock front seen in the MHD benchmark/$\lambda_i=0.0072$ cases are no longer present, leaving most of the remaining fast wave concentrated in four perpendicular shock fronts. The dispersive whistler waves cause the field near the null to rapidly oscillate, even before the arrival of the shock fronts. As the waves from different quadrants arrive out of phase with one another, multiple null pairs are generated near the equilibrium X-point location, prior to the arrival of a pair of perpendicular magnetic shocks at the null. The initial shocks set up a current sheet containing an island of current surrounding an O-type null at the centre, with up to 6 additional nulls at either side. A second slower pair of perpendicular shocks (arriving from the opposite quadrants to the first) destabilise the current sheet, causing all additional nulls to recombine, leaving a single X-type null remaining at the origin.

 Following the initial X-point collapse, all simulations undergo oscillatory reconnection, through which the system is able to relax. Both in MHD and Hall MHD, plasma pressure is enhanced at both ends of the initial current sheet, which exerts a force to open the magnetic field lines in the compressed quadrants, in an attempt to restore equilibrium. As the field nears its original configuration, both the gas pressure and Lorentz forces balance each other, but plasma inertia causes the field to overshoot, forming a second (perpendicular) current sheet. A Lorentz force response accelerates plasma back towards its equilibrium configuration once more, and the cycle repeats; each additional overshoot is smaller than the last, and requires a smaller restoration force as the system returns to equilibrium.
%

The simulation results presented in this paper indicate that two-dimensional X-point reconnection is rather more complicated in Hall MHD than in MHD. In the former case, many more magnetic nulls are created (albeit transiently) and the presence of whistler waves means that information can propagate more rapidly than in the MHD limit. Despite these significant differences, it is striking that over long timescales both the rate of oscillatory reconnection and the damping rate of this oscillation are essentially invariant under changes in the normalised ion skin depth, $\lambda_i$ (see Fig.~\ref{fig:cnullevol}). As noted in Sect.~\ref{sec:Setup}, for a given set of values of $B_0$, $n_0$ and $T_0$, increasing the value of $\lambda_i$ is equivalent to reducing the normalising length scale $l_0$ and hence the Alfv\'en time $t_0 = l_0/c_A$. Thus, if we also fix the dimensionless resistivity $\eta$, the absolute timescales of the reconnection process indicated by Fig.~\ref{fig:cnullevol} become progressively shorter as the normalised ion skin depth is increased. It should be noted that holding $\eta$ fixed, as we have done in these simulations, while reducing $l_0$ also implies a reduction in the absolute resistivity, $\eta_0$ (or, equivalently, a smaller enhancement of the resistivity above the collisional value). Thus, Fig.~\ref{fig:cnullevol} reinforces the conclusion of previous authors \citep[e.g.][]{paper:Birnetal2001} that the Hall term produces an enhanced rate of reconnection.

Even for the highest value of $\lambda_i$ used in the simulations ($0.072$), $l_0$ is still more than an order of magnitude larger than $\delta_i$, although for typical coronal parameters this is well below the spatial resolution of any measurements. Whether magnetic structures with sub-resolution macroscopic length scales of this order are actually present in the corona is unclear. If they are, Fig.~\ref{fig:cnullevol} indicates that the timescale of reconnection within these structures could be very rapid. For example, with $B_0 = 100\,$G, $n_0 = 10^{16}$m$^{-3}$ and $\lambda_i = 0.072$, we find that $t_0 \simeq 16\mu s$. The damping time of the long timescale oscillation in Fig.~\ref{fig:cnullevol} is then no more than 50$t_0 \simeq 1\,$ms. The most rapid fluctuations reported in flare hard X-ray emission occurred on timescales that were more than an order of magnitude longer than this 19098responseome of the most rapid fluctuations have been observed in flare hard X-ray emission, occurring on timescales that are more than an order of magnitude longer than this
 \citep{paper:Kiplingeretal1983,paper:Dmitrievetal2006}. Thus, taking into account the Hall term in the generalised Ohm's law makes it possible to obtain reconnection rates that are at least as rapid as those implied by flare observations.

Several extensions to this work merit further consideration. For example, introducing an initially non-zero plasma beta may significantly alter the balance of forces in the oscillatory cycle, potentially affecting the oscillation frequency and hence the rate of reconnection during the oscillatory phase. This would also introduce slow magnetoacoustic waves into the problem; in Hall MHD, the additional presence of coupled fast and shear {\A} wave motions would further complicate any identification of specific wave modes. Altering the form of the initial disturbance (even simply in MHD) would alter the wave-speed asymmetry and hence affect the arrival of the initial perpendicular shock-waves at the null. Finally, full 3D Hall MHD studies of separator/null-point reconnection (for example) would allow a more comprehensive comparison between the full vector forms of the magnetic vector potential, ${\bf{A}}$, and current ${\bf{J}}$. Such a comparison may provide further insight into the influence of planar components of current (generated by the Hall term) on the reconnection rate.

\begin{acknowledgements}
This work was part-funded by the RCUK Energy Programme under grant EP/I501045 and the United Kingdom Engineering and Physical Sciences Research Council through a CASE studentship. JT would like to thank T. Neukirch (University of St Andrews) and J. McLaughlin (University of Northumbria) for helpful discussions. IDM acknowledges support of a Royal Society University Research Fellowship.
\end{acknowledgements}

\bibliographystyle{aa}        
\bibliography{19098}          

\begin{thebibliography}{39}
\expandafter\ifx\csname natexlab\endcsname\relax\def\natexlab#1{#1}\fi

\bibitem[{{Antiochos} {et~al.}(1999){Antiochos}, {DeVore}, \&
  {Klimchuk}}]{paper:Antiochosetal1999}
{Antiochos}, S.~K., {DeVore}, C.~R., \& {Klimchuk}, J.~A. 1999, \apj, 510, 485

\bibitem[{{Arber} {et~al.}(2001){Arber}, {Longbottom}, {Gerrard}, \&
  {Milne}}]{paper:LareXd2001}
{Arber}, T.~D., {Longbottom}, A.~W., {Gerrard}, C.~L., \& {Milne}, A.~M. 2001,
  Journal of Computational Physics, 171, 151

\bibitem[{{Birn} {et~al.}(2001){Birn}, {Drake}, {Shay}, {Rogers}, {Denton},
  {Hesse}, {Kuznetsova}, {Ma}, {Bhattacharjee}, {Otto}, \&
  {Pritchett}}]{paper:Birnetal2001}
{Birn}, J., {Drake}, J.~F., {Shay}, M.~A., {et~al.} 2001, \jgr, 106, 3715

\bibitem[{{Birn} \& {Priest}(2007)}]{book:BirnPriest}
{Birn}, J. \& {Priest}, E. 2007, Reconnection of Magnetic Fields (New York:
  Cambridge University Press), 94--95

\bibitem[{{Biskamp}(2000)}]{book:Biskamp}
{Biskamp}, D. 2000, {Magnetic Reconnection in Plasmas} (Cambridge University
  Press)

\bibitem[{{Craig} \& {Litvinenko}(2002)}]{paper:CraigLitvinenko2002}
{Craig}, I.~J.~D. \& {Litvinenko}, Y.~E. 2002, \apj, 570, 387

\bibitem[{{Craig} \& {Litvinenko}(2008)}]{paper:CraigLitvinenko2008}
{Craig}, I.~J.~D. \& {Litvinenko}, Y.~E. 2008, \aap, 484, 847

\bibitem[{{Craig} \& {McClymont}(1991)}]{paper:CraigMcClymont1991}
{Craig}, I.~J.~D. \& {McClymont}, A.~N. 1991, \apjl, 371, L41

\bibitem[{{Craig} \& {McClymont}(1993)}]{paper:CraigMcClymont1993}
{Craig}, I.~J.~D. \& {McClymont}, A.~N. 1993, \apj, 405, 207

\bibitem[{{Craig} \& {Watson}(1992)}]{paper:CraigWatson1992}
{Craig}, I.~J.~D. \& {Watson}, P.~G. 1992, \apj, 393, 385

\bibitem[{{Craig} \& {Watson}(2003)}]{paper:CraigWatson2003}
{Craig}, I.~J.~D. \& {Watson}, P.~G. 2003, \solphys, 214, 131

\bibitem[{{De Moortel}(2005)}]{review:DeMoortel2005}
{De Moortel}, I. 2005, Royal Society of London Philosophical Transactions
  Series A, 363, 2743

\bibitem[{{De Moortel} \& {Nakariakov}(2012)}]{review:DeMoortelNakariakov2012}
{De Moortel}, I. \& {Nakariakov}, V.~M. 2012, Royal Society of London
  Philosophical Transactions Series A (submitted), arXiv:1202.1944
  [astro-ph.SR]

\bibitem[{{Dmitriev} {et~al.}(2006){Dmitriev}, {Kudryavtsev}, {Lazutkov},
  {Matveev}, {Savchenko}, {Skorodumov}, \& {Charikov}}]{paper:Dmitrievetal2006}
{Dmitriev}, P.~B., {Kudryavtsev}, I.~V., {Lazutkov}, V.~P., {et~al.} 2006,
  Solar System Research, 40, 142

\bibitem[{{Dorelli}(2003)}]{paper:Dorelli2003}
{Dorelli}, J.~C. 2003, Physics of Plasmas, 10, 3309

\bibitem[{Dungey(1953)}]{paper:Dungey1953}
Dungey, J. 1953, Philosophical Magazine Series 7, 44, 725

\bibitem[{{Fletcher} {et~al.}(2007){Fletcher}, {Hannah}, {Hudson}, \&
  {Metcalf}}]{paper:Fletcheretal2007}
{Fletcher}, L., {Hannah}, I.~G., {Hudson}, H.~S., \& {Metcalf}, T.~R. 2007,
  \apj, 656, 1187

\bibitem[{{Hassam}(1992)}]{paper:Hassam1992}
{Hassam}, A.~B. 1992, \apj, 399, 159

\bibitem[{{Haynes} \& {Parnell}(2007)}]{paper:HaynesParnell2007}
{Haynes}, A.~L. \& {Parnell}, C.~E. 2007, Physics of Plasmas, 14, 082107

\bibitem[{{Hesse} \& {Schindler}(1988)}]{paper:HesseSchindler1988}
{Hesse}, M. \& {Schindler}, K. 1988, \jgr, 93, 5559

\bibitem[{{Kiplinger} {et~al.}(1983){Kiplinger}, {Dennis}, {Frost}, {Orwig}, \&
  {Emslie}}]{paper:Kiplingeretal1983}
{Kiplinger}, A.~L., {Dennis}, B.~R., {Frost}, K.~J., {Orwig}, L.~E., \&
  {Emslie}, A.~G. 1983, \apjl, 265, L99

\bibitem[{{Longcope} \& {Parnell}(2009)}]{paper:LongcopeParnell2009}
{Longcope}, D.~W. \& {Parnell}, C.~E. 2009, \solphys, 254, 51

\bibitem[{{Masson} {et~al.}(2012){Masson}, {Aulanier}, {Pariat}, \&
  {Klein}}]{paper:Massonetal2012}
{Masson}, S., {Aulanier}, G., {Pariat}, E., \& {Klein}, K.-L. 2012, \solphys,
  276, 199

\bibitem[{{Masson} {et~al.}(2009){Masson}, {Pariat}, {Aulanier}, \&
  {Schrijver}}]{paper:Massonetal2009}
{Masson}, S., {Pariat}, E., {Aulanier}, G., \& {Schrijver}, C.~J. 2009, \apj,
  700, 559

\bibitem[{{McClements} {et~al.}(2004){McClements}, {Thyagaraja}, {Ben Ayed}, \&
  {Fletcher}}]{paper:McClementsetal2004}
{McClements}, K.~G., {Thyagaraja}, A., {Ben Ayed}, N., \& {Fletcher}, L. 2004,
  \apj, 609, 423

\bibitem[{{McLaughlin} {et~al.}(2009){McLaughlin}, {De Moortel}, {Hood}, \&
  {Brady}}]{paper:McLaughlinetal2009}
{McLaughlin}, J.~A., {De Moortel}, I., {Hood}, A.~W., \& {Brady}, C.~S. 2009,
  \aap, 493, 227

\bibitem[{{McLaughlin} \& {Hood}(2004)}]{paper:McLaughlinHood2004}
{McLaughlin}, J.~A. \& {Hood}, A.~W. 2004, \aap, 420, 1129

\bibitem[{{McLaughlin} \& {Hood}(2005)}]{paper:McLaughlinHood2005}
{McLaughlin}, J.~A. \& {Hood}, A.~W. 2005, \aap, 435, 313

\bibitem[{{McLaughlin} \& {Hood}(2006)}]{paper:McLaughlinHood2006b}
{McLaughlin}, J.~A. \& {Hood}, A.~W. 2006, \aap, 459, 641

\bibitem[{{McLaughlin} {et~al.}(2011){McLaughlin}, {Hood}, \& {de
  Moortel}}]{review:McLaughlinetal2011}
{McLaughlin}, J.~A., {Hood}, A.~W., \& {de Moortel}, I. 2011, \ssr, 158, 205

\bibitem[{{Nakariakov} \& {Verwichte}(2005)}]{review:NakariakovVerwichte2005}
{Nakariakov}, V.~M. \& {Verwichte}, E. 2005, Living Reviews in Solar Physics,
  2, 3

\bibitem[{{Ofman} {et~al.}(1993){Ofman}, {Morrison}, \&
  {Steinolfson}}]{paper:Ofmanetal1993}
{Ofman}, L., {Morrison}, P.~J., \& {Steinolfson}, R.~S. 1993, \apj, 417, 748

\bibitem[{{Pariat} {et~al.}(2009){Pariat}, {Antiochos}, \&
  {DeVore}}]{paper:Pariatetal2009}
{Pariat}, E., {Antiochos}, S.~K., \& {DeVore}, C.~R. 2009, \apj, 691, 61

\bibitem[{{Parker}(1957)}]{paper:Parker1957}
{Parker}, E.~N. 1957, \jgr, 62, 509

\bibitem[{{Priest} \& {Forbes}(2007)}]{book:PriestForbes}
{Priest}, E. \& {Forbes}, T. 2007, Magnetic Reconnection: MHD Theory and
  Applications (Cambridge University Press)

\bibitem[{{Schindler} {et~al.}(1988){Schindler}, {Hesse}, \&
  {Birn}}]{paper:Schindleretal1988}
{Schindler}, K., {Hesse}, M., \& {Birn}, J. 1988, \jgr, 93, 5547

\bibitem[{{Senanayake} \& {Craig}(2006)}]{paper:SenanayakeCraig2006}
{Senanayake}, T. \& {Craig}, I.~J.~D. 2006, \aap, 451, 1117

\bibitem[{{Sweet}(1958)}]{paper:Sweet1958}
{Sweet}, P.~A. 1958, in IAU Symposium, Vol.~6, Electromagnetic Phenomena in
  Cosmical Physics, ed. {B.~Lehnert}, 123

\bibitem[{{Threlfall} {et~al.}(2011){Threlfall}, {McClements}, \& {de
  Moortel}}]{paper:Threlfalletal2011}
{Threlfall}, J., {McClements}, K.~G., \& {de Moortel}, I. 2011, \aap, 525, A155

\end{thebibliography}
\end{document}